\documentclass[showpacs,preprintnumbers,amsmath,amssymb,floatfix]{revtex4}

\usepackage{graphicx}
\usepackage{dcolumn}
\usepackage{bm}
\newcommand{\ud}{\mathrm{d}}

\newcommand{\ui}{\mathrm{i}}
\newcommand{\nn}{\nonumber}
\newcommand{\hp}{\#_\perp}

\newcommand{\cp}{\delta_\perp}

\newcommand{\vt}{\widetilde{V}}

\newcommand{\bp}{\beta}
\newcommand{\dpe}{\ud_\perp}
\newcommand{\doma}{\mathcal{D}}
\newcommand{\domb}{{\partial\mathcal{D}}}
\newcommand{\matf}{\mathcal{F}}
\newcommand{\hr}{\mathcal{H}}

\newcommand{\rtz}{(r,\theta,\zeta)}
\newcommand{\rtu}{(r,\theta,u)}
\newcommand{\hW}{\hat{W}}
\newcommand{\hX}{\hat{X}}
\newcommand{\cX}{{\check X}{}}
\newcommand{\chr}{\check{\mathcal{H}}{}}
\newcommand{\cW}{{\check {W}}{}}
\newcommand{\intfin}{\int_{-\infty}^{\infty}}

\newcommand{\Rep}{\check{R}}

\newcommand{\Rhat}{\check{R}}

\begin{document}


\title{Wake potentials and impedances of charged beams in gradually
  tapering structures}

\author{D.A. Burton}
\author{D.C. Christie}
\author{J.D.A. Smith}
\author{R.W. Tucker}
\affiliation{%
Department of Physics, Lancaster University, LA1 4YB, UK\\ \& The Cockcroft
Institute, Daresbury Science and Innovation Campus, Daresbury, Warrington WA4 4AD, UK
}%

\date{\today}

\begin{abstract}
We develop an analytical method for calculating the geometric wakefield and
impedances of an ultrarelativistic beam propagating on- and off-axis
through an axially symmetric geometry with slowly varying circular cross-section, such as a
transition. Unlike previous analytical methods, our approach
permits detailed perturbative investigation of geometric wakefields,
and detailed perturbative investigation of impedance as a function of frequency.    
We compare the accuracy of the results of our approach with numerical
simulations performed using the code ECHO and determine parameters in which there is
good agreement with our asymptotic analysis.
\end{abstract}


\maketitle
\section{Introduction}
\label{ch_intro}
Vital considerations in modern accelerator and light source design
include the influence of non-uniform metallic structures, such as a vacuum system or collimator jaws, on nearby charged particle beams.
Rapid changes in the spatial profile of the structure tend to have
undesirable consequences for a particle beam, such as inducing
instabilities and emittance growth; hence, designers often employ
structures with cross-sections that gradually vary with distance. For
example, a series of
gradually tapering metallic structures (collimators) may be used to strip
unwanted particles from beams prior to the collision event (see, for
example,~\cite{smith_glasman07}).      

Optimal design of particle accelerator subsystems is commonly sought
by direct numerical solution of Maxwell's equations, but
accurate numerical computation of the electromagnetic fields near
a short bunch in subsystems such as a post-linac
collimator requires large computing resources~\cite{beard_smith06}.
In particular, numerically
resolving a collimator with a gradually tapering geometry requires a
mesh whose cells are much shorter than the length of the
collimator. Although one may employ windowing techniques to avoid calculating
the fields throughout the entire structure at every time step, such
calculations frequently require intensive parallel computation
~\cite{gjonaj06, smith09}.
Similar problems are encountered when
considering beam pipe transitions inside small-gap undulators in
future light sources. In such cases, analytical expressions for the beam's
behaviour are highly desirable.

In practice, it is assumed that the beam does not deviate much from
rectilinear motion parallel to the axis of the metallic structure~\cite{chaobk}.
The beam's trajectory is often obtained as a perturbation due to the
coupling impedance of the unperturbed beam and structure (early discussions of
impedances in tapered structures were given by Yokoya~\cite{yokoya90}
and Stupakov~\cite{stupakov96}). Although each frequency
component of the unperturbed beam leads to its own impedance, it has
been noted that low-frequency transverse impedance
is dominated by the zero-frequency component of the source and longitudinal
impedance is directly proportional to
frequency over a broad frequency range~\cite{stupakov96}. Recent
work~\cite{podobedov06, stupakov07} has been tailored
to the above observations; it yields a frequency independent result
for the transverse impedance and a longitudinal
impedance directly proportional to frequency. Such methods do not permit detailed exploration of
impedances as a function of frequency; the purpose of the following is
to address this limitation.

To overcome the above, a new scheme for obtaining
analytical expressions of impedances in gradually tapering, axially
symmetric structures was suggested by us in~\cite{burton08}.
In common with~\cite{podobedov06, stupakov07}, our
approach employs an expansion in a small parameter $\epsilon$
characterising the gradually changing cross-sectional radius of the structure. However,
we employ a decomposition of Maxwell's equations using auxiliary
potentials that yields impedance as a series in frequency, without a
priori assuming that the frequency is small.

The following is an extensive investigation of the approach introduced
in~\cite{burton08}. We study the longitudinal and transverse wake
potentials of
a bunch travelling parallel to the axis of a perfectly
conducting axially symmetric structure and compare the second, fourth and sixth order results in
$\epsilon$ with numerical data from ECHO~\cite{zagorodnov05}. We then
present analytical expressions for longitudinal impedance; our method
applied to a harmonic source reproduces the results given by
Yokoya~\cite{yokoya90} and Stupakov~\cite{stupakov96, stupakov07} when
working to second order in $\epsilon$. We also present corrections to
the Yokoya-Stupakov results that arise due to higher order terms in
$\epsilon$. Expressions for the longitudinal impedance up to
fourth order in $\epsilon$ have been given previously
in~\cite{burton08} and we include them here for completeness.
We then turn to a detailed exposition of the passage from Maxwell's equations to
our iterative procedure for calculating auxiliary potentials order-by-order in
$\epsilon$. A discussion of the difficulties encountered when
auxiliary potentials are not used is given in Appendix~\ref{juste}.
\section{Overview and results}
\label{section:overview_and_results}
A full account of our solution method is presented in later
sections. This section focusses on a comparison of the results
obtained using our asymptotic method and results obtained using the code ECHO.
 
\begin{figure}
\centering
\scalebox{0.5}{\includegraphics[angle=90,width=\textwidth]{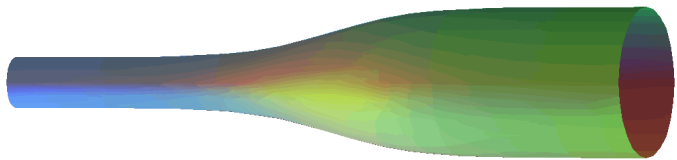}}
\caption{A waveguide with a slowly varying profile\label{schematic_waveguide}}
\end{figure}

Our investigation concentrates on beams propagating through axially symmetric
structures whose circular cross-sections gradually change along the axis of symmetry. Such structures are topologically equivalent to an infinite right
circular cylinder and we call them \emph{waveguides} (see
Figure~\ref{schematic_waveguide}). This article focusses entirely on
the \emph{geometric} wake of the beam in the inductive
regime~\cite{tenenbaum:2007}. Here, the walls of the waveguide
are assumed to be perfectly conducting; resistive effects will be
considered elsewhere.

We develop the electromagnetic field of the
beam as an asymptotic expansion in a small parameter $\epsilon$
characterising the gradually changing radius of a waveguide with a
smoothly varying profile.
In particular, the radial profile of the waveguide is specified as
\begin{equation}
r = R(z) = \Rhat(\epsilon z)
\end{equation}
where $\epsilon \ll 1$ and $(r,\theta,z)$ is a cylindrical coordinate
system with $z$ a Cartesian coordinate parallel to the waveguide's
axis of symmetry ($r=0$).

The two radial profiles explored here numerically are
\begin{equation}
\label{exp_profile}
r = R(z) = 20-18 \exp[-z^2/(8\times 10^5 l^2)]
\end{equation}
with $\epsilon=1/\sqrt{8\times 10^5}$ and
\begin{equation}
\label{sech_profile}
r = R(z) = 20 - 18 \textrm{ sech } (0.01 z/l)
\end{equation}
where $\epsilon = 0.01$. The coordinates $z,r$ and the constant $l$ have
 dimensions of length; in the following $l=1\text{mm}$, and $z, r$ are measured in
 $\text{mm}$. The profiles (\ref{exp_profile}, \ref{sech_profile})
have a ratio of gap to beam pipe radius and average taper gradient similar to that proposed
for the beam delivery system of the next generation of lepton
colliders~\cite{tenenbaum:2007}.

Wake potentials are calculated by integrating
components of the electromagnetic field along a straight line parallel to the
axis of the waveguide. We consider fields excited by a narrow Gaussian bunch of effective
length $\sigma$ propagating at the speed of light parallel to the
waveguide's axis, offset by a distance $r_0$ from the waveguide's
axis. The charge density $\rho$ of the bunch is  
\begin{equation}
\label{original_source}
\rho_{r_0,\theta_0}(r,\theta,u) = \frac{q_0e^{-\frac{u^2}{2\sigma^2}}}{\sigma\sqrt{2\pi}}\delta\left(x-r_0 \cos \theta_0\right)\delta\left(y-r_0\sin\theta_0\right)
\end{equation}
where $u=z-ct$, $x=r\cos\theta$, $y=r\sin\theta$ and $q_0$ is the
total charge of the bunch. The longitudinal wake potential $W^\parallel$ is given by
\begin{equation}
W^\parallel\rtu = -\frac{1}{q_0}\int_{-\infty}^\infty
E_z\big|_{ct=z-u}\,dz
\end{equation}
where
\begin{equation}
E_z\big|_{ct=z-u} = E_z\left(r,\theta,z,t=\frac{z-u}{c}\right)
\end{equation}
and $E_z$ is the longitudinal (i.e. $z$) component of the electric
field. The transverse wake potential ${\bf W}_\perp$ may be
obtained from $W^\parallel$ using the Panofsky-Wenzel
relation~\cite{panofsky56}.

\begin{figure}
\centering
\scalebox{0.75}{\includegraphics[width=1.0\textwidth]{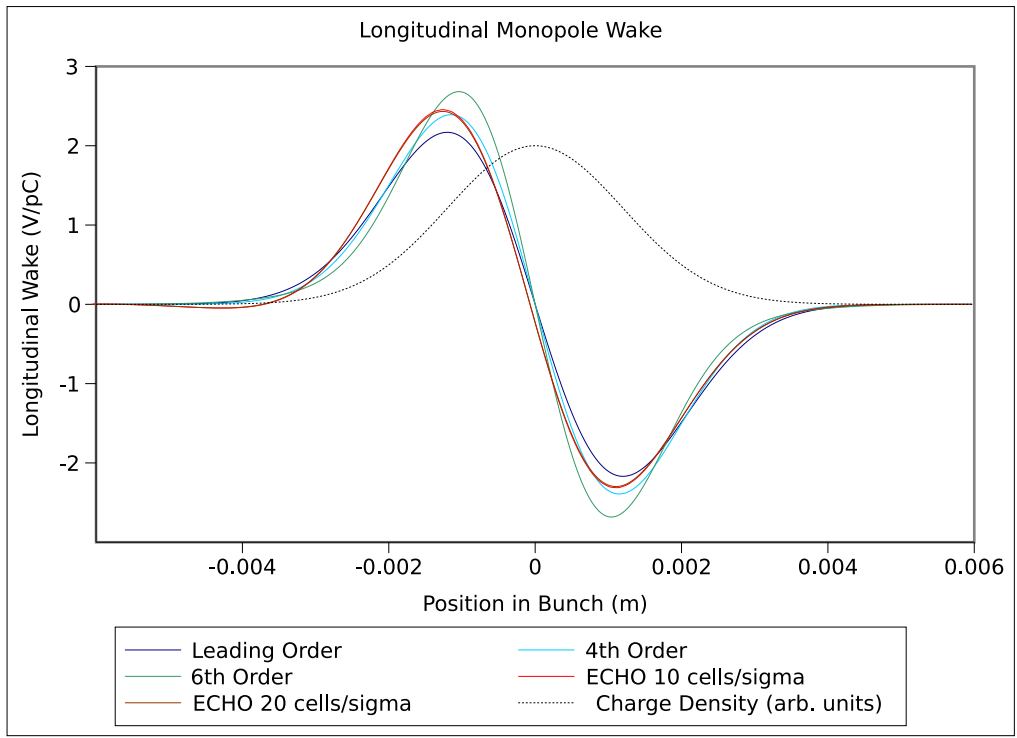}}
\caption{Sample longitudinal wake potential $W^\parallel$ versus
  $u=z-ct$ for an on-axis source (length $1.2\text{mm}$) propagating through the exponential
  geometry (\ref{exp_profile}).
\label{long_wake_versus_position_exp}}
\end{figure}

\begin{figure}
\centering
\scalebox{0.75}{\includegraphics[width=1.0\textwidth]{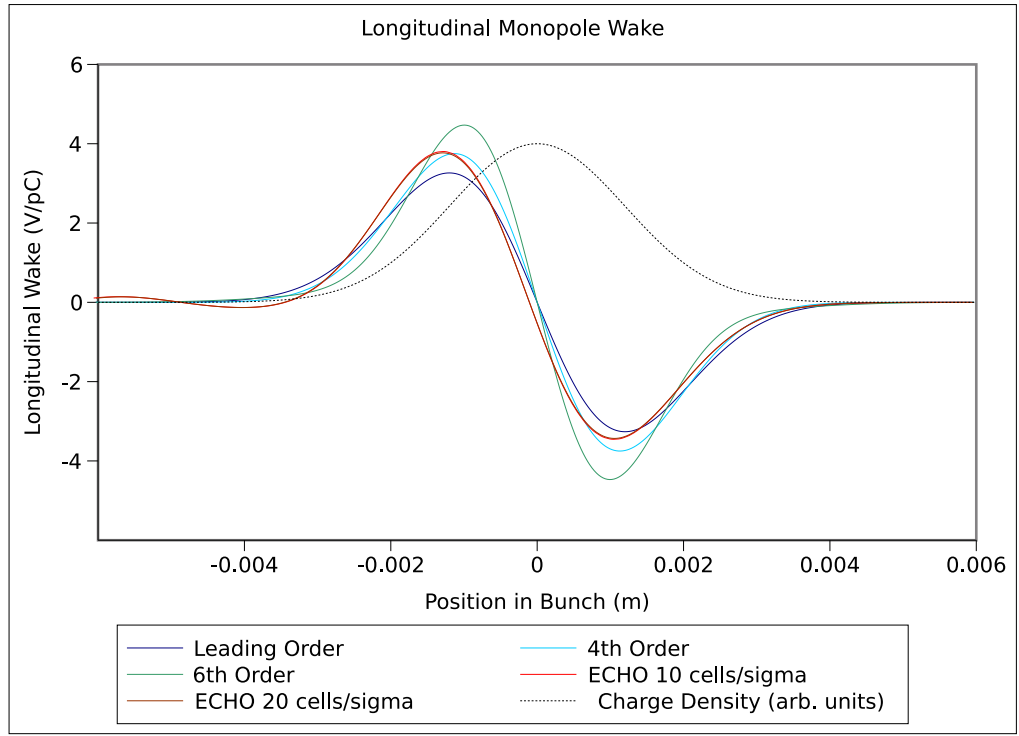}}
\caption{Sample longitudinal wake potential $W^\parallel$ versus
  $u=z-ct$ for an on-axis source (length $1.2\text{mm}$) propagating
  through the hyperbolic secant geometry
  (\ref{sech_profile}).
\label{long_wake_versus_position_sech}}
\end{figure}

Sample results for $W^\parallel$ are shown in Figures
\ref{long_wake_versus_position_exp},
\ref{long_wake_versus_position_sech}. The ``leading order'' curves result
from electromagnetic fields calculated up to 2nd order in $\epsilon$,
while the ``4th order'' and ``6th order'' curves arise from including 4th order
and 6th order corrections in $\epsilon$ respectively. It may be shown
that the 3rd, 5th and 7th order contributions to $W^\parallel$ vanish.

Wake potentials can be calculated from
impedances using Fourier methods~\cite{zotter_kheifets98} and the
leading order results may be recovered using well-known expressions
for impedance originally due to Yokoya~\cite{yokoya90}. 
Corrections to the leading order transverse impedance that properly
accommodate the waveguide's taper but neglect the frequency
dependence of the source may be found
in~\cite{podobedov06}; we expect the frequency dependences of
impedances to be more accurately represented using our method.

Data from ECHO was obtained for a sequence of mesh cell densities ($10,
15, 20$ cells over bunch length $\sigma$; we judged the numerical
errors to be small based on the small differences between the results for
$10$, $15$ and $20$ cells per $\sigma$). The monopole
and dipole contributions to (\ref{original_source}) are the zeroth and
first order terms in an expansion of (\ref{original_source}) with
respect to $r_0$; for direct comparison with the ECHO data, only the
wake potentials due the monopole and dipole contributions are calculated.  

\begin{figure}
\centering
\scalebox{0.75}{\includegraphics[width=\textwidth]{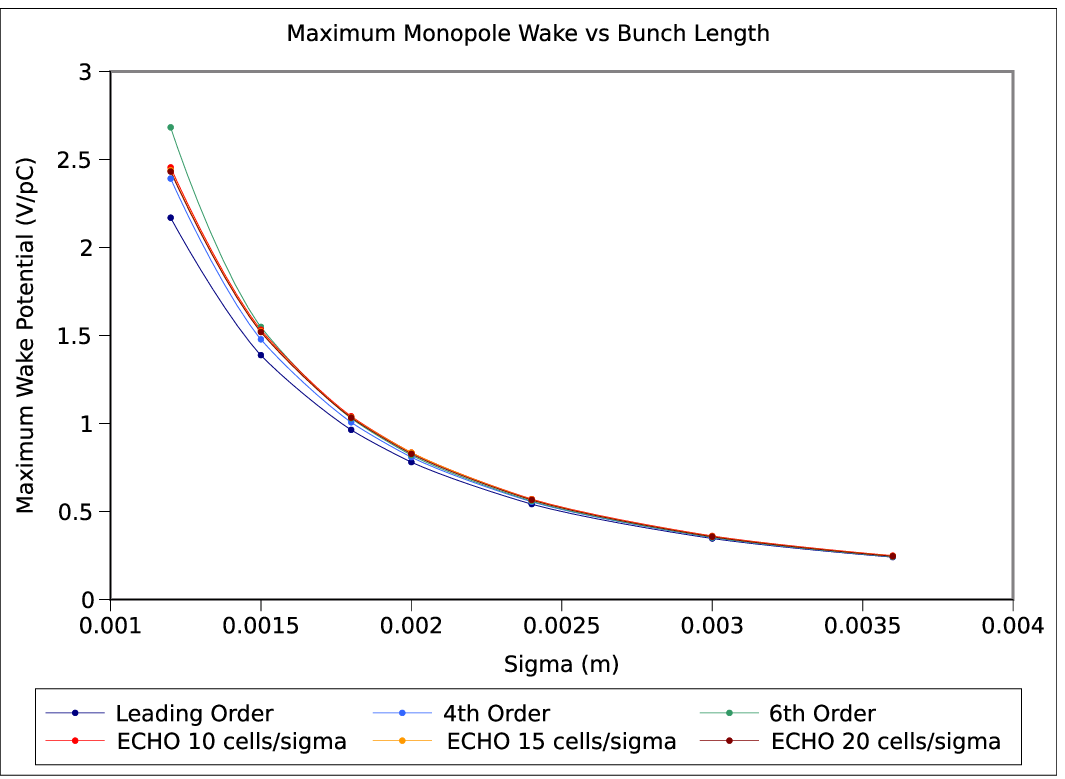}}
\caption{\label{fig:first_exp} The maximum in $u$ of $W^\parallel$ versus effective bunch length
  $\sigma$ for a monopole source propagating through the exponential geometry (\ref{exp_profile}).}
\end{figure}

\begin{figure}
\centering
\scalebox{0.75}{\includegraphics[width=\textwidth]{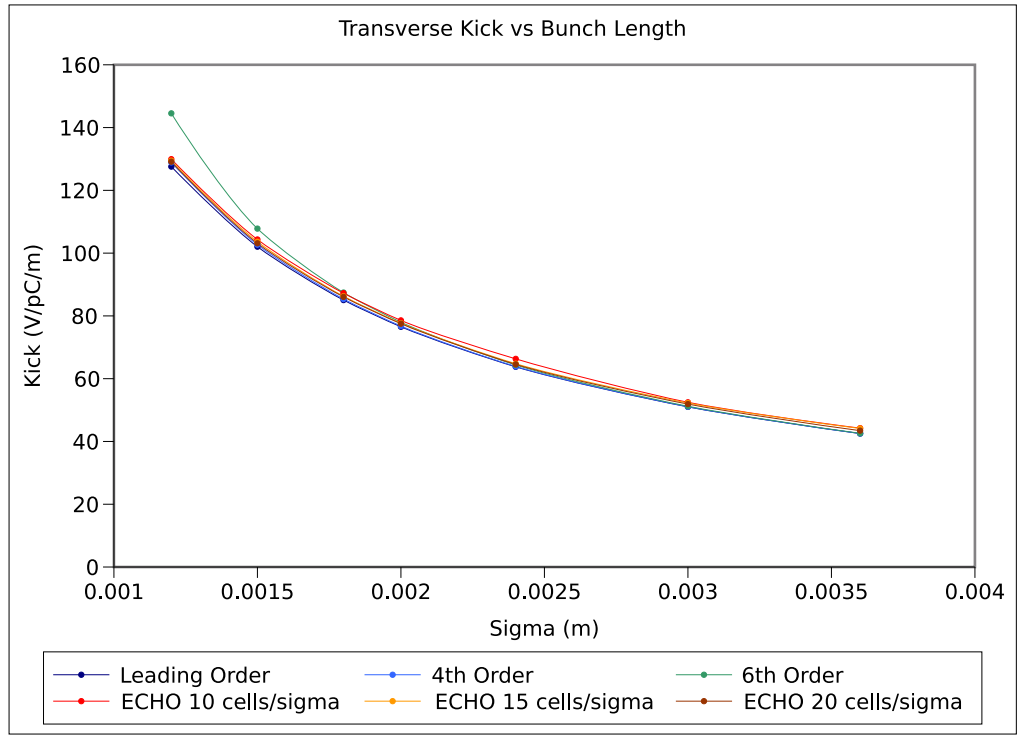}}
\caption{\label{fig:last_exp} The transverse momentum kick
  versus effective bunch length $\sigma$ for a dipole source
  propagating through the exponential geometry
  (\ref{exp_profile}).}
\end{figure}

\begin{figure}
\centering
\scalebox{0.75}{\includegraphics[width=\textwidth]{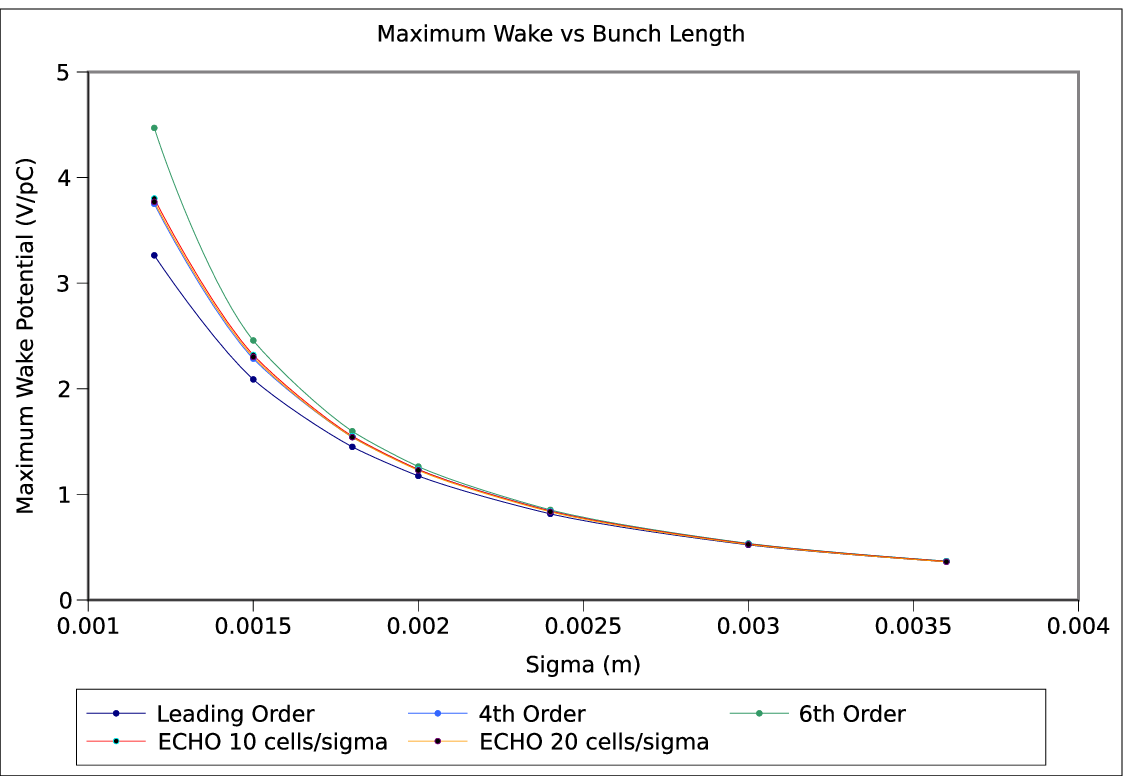}}
\caption{\label{fig:first_sech} The maximum in $u$ of $W^\parallel$
  versus effective bunch length $\sigma$ for a monopole source
  propagating through the hyperbolic secant geometry (\ref{sech_profile}).}
\end{figure}

\begin{figure}
\centering
\scalebox{0.75}{\includegraphics[width=\textwidth]{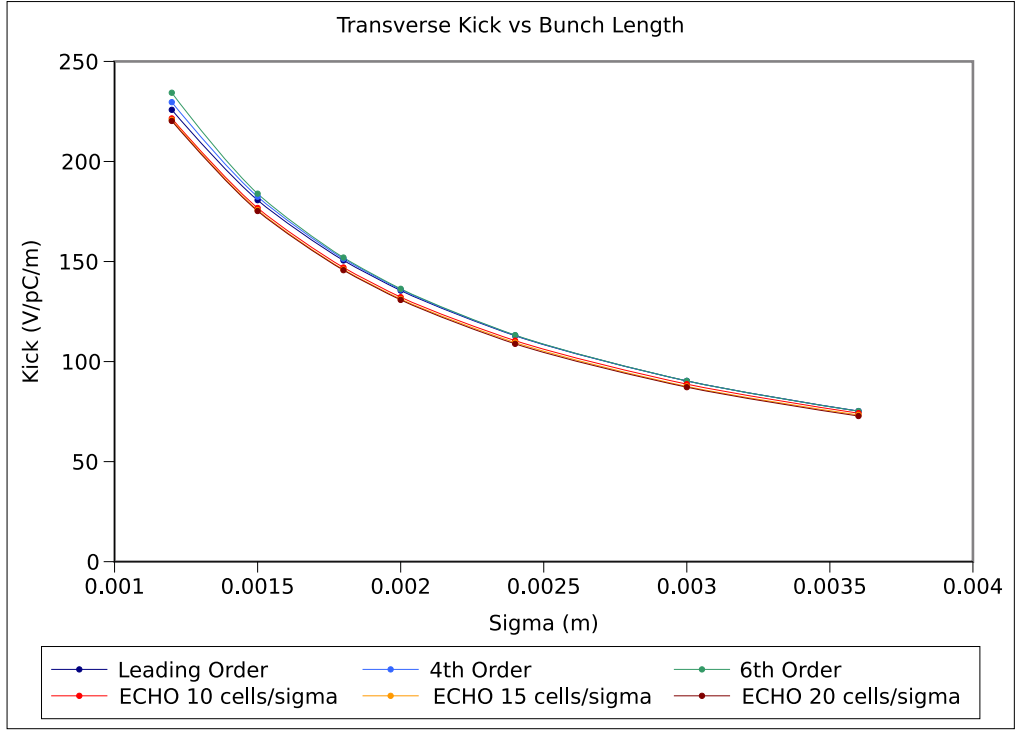}}
\caption{\label{fig:last_sech} The transverse
  momentum kick versus effective bunch length $\sigma$ for a dipole
  source propagating through the hyperbolic secant geometry
  (\ref{sech_profile}).}
\end{figure}

Figures (\ref{fig:first_exp}-\ref{fig:last_exp}) show the magnitude
and position of the maxima of
longitudinal wake potentials, the magnitude of the maxima of the radial component of
transverse wake potentials and the momentum kicks induced by
transverse wakes in the exponential geometry. Figures
(\ref{fig:first_sech}-\ref{fig:last_sech}) show the corresponding
quantities for the sech profile. The higher order contributions
reduce in significance as 
the effective bunch length $\sigma$ increases and we observe that the
4th and 6th order results agree well with the leading order
predictions for bunch lengths longer than $\sim
2\text{mm}$. For $\sigma$ less than $\sim 1.5\text{mm}$ the data shows
that, in a number of cases, the maximum
values of the wake potentials and momentum kicks obtained using ECHO
generally agree better with the 4th order results than with the leading order
(Yokoya-Stupakov) predictions over the parameter ranges considered
here. The sample plots of $W^\parallel$ shown in Figures
\ref{long_wake_versus_position_exp},
\ref{long_wake_versus_position_sech} demonstrate this for a bunch
length of $1.2\text{mm}$; for such bunch lengths the reliability
of the 6th order correction is questionable. Although the agreement with the numerical results is
impressive, we warn the reader that the radius of convergence of our
perturbative expansion is unknown.
\subsection{Impedance of an on-axis beam}
\label{section:impedance_on-axis_beam}
In this section we present impedance formulae due to the on-axis harmonic charge
density
\begin{equation}
\label{on-axis_line_source}
\varrho_\omega(r,\theta,u) = \frac{I_\omega}{c} e^\frac{i\omega u}{c}\delta(x)\delta(y)
\end{equation}  
where $I_\omega$ defines the harmonic current component of the beam.
The longitudinal impedance $Z^\parallel(\omega)$ for angular frequency
$\omega$ is
\begin{equation}\label{zintone}
Z^\parallel(\omega)=-\frac{1}{I_\omega}\int_{-\infty}^\infty{e^{-\frac{i\omega u}{c}}
  E_z\big|_{ct=z-u}\ \ud z}
\end{equation}
where $E_z$ is the longitudinal component of the electric field
generated by the source (\ref{on-axis_line_source}). The approach detailed in
subsequent sections leads to the following :
\begin{align}
\label{monopole_Z}
Z^\parallel_{\text{on-axis}}(\omega) = Z^\parallel_{1\,\text{on-axis}} +
Z^\parallel_{2\,\text{on-axis}} + Z^\parallel_{4\,\text{on-axis}}
+ Z^\parallel_{6\,\text{on-axis}} + \dots
\end{align}
where $Z^\parallel_{1\,\text{on-axis}} + Z^\parallel_{2\,\text{on-axis}}$ is the
Yokoya-Stupakov longitudinal geometric impedance
\begin{align}
\label{imp_Z0}
&Z^\parallel_{1\,\text{on-axis}} =
\frac{1}{2\pi\varepsilon_0 c}\ln\left(\frac{R_1}{R_2}\right),\\
\label{imp_Z2}
&Z^\parallel_{2\,\text{on-axis}} = -\frac{i\omega}{4\pi\varepsilon_0 c^2} \int_{-\infty}^{\infty}{R'^2\ud
 z}
\end{align}
with $R'=dR/dz$, $R_1 = R(\infty)$
and $R_2 = R(-\infty)$. Equations (\ref{imp_Z0}, \ref{imp_Z2}) follow,
respectively, from 1st and 2nd order terms in $\epsilon$
in an asymptotic series for $E_z$. The corrections
\begin{align}
\label{imp_Z4}
4\pi\varepsilon_0 c Z^\parallel_{4\,\text{on-axis}} = \frac{i\omega}{24 c} \int_{-\infty}^{\infty}
\bigg\{ 5 R'^4+3 \left(R\,R''\right)^2 -
2\frac{\omega^2}{c^2}\left(R^2\,R''\right)^2  \bigg\} \ud z
\end{align}
and
\begin{align}
\notag
4 \pi \varepsilon_0 c Z^\parallel_{6\,\text{on-axis}} =&
-\frac{i\omega}{c}\int_{-\infty}^{\infty}\bigg\{ \frac{3}{16}(R''
 \,R'\,R)^2
+ \frac{11}{120} R'^6 +\frac{1}{48}(R^2\,R''')^2\bigg\}\ud z\nn\\
&+  \frac{i\omega^3}{c^3}\int_{-\infty}^{\infty}\bigg\{\frac{11}{256}
  \left(R^3\,R'''\right)^2 -\frac{1}{6}
  \left(R^2\,R'\,R''\right)^2\nn
- \frac{73}{768}R^5\,R''^3\bigg\}\ud z\nn\\
\label{imp_Z6} 
&+ \frac{i\omega^5}{c^5}\int_{-\infty}^{\infty}\bigg\{ \frac{19}{160} (R^3\,R'\,R'')^2
+ \frac{73}{1920} R^7\,R''^3
- \frac{19}{1920}(R'''\,R^4)^2\bigg\}\ud z
\end{align}
arise from 4th and 6th order terms in the asymptotic
series for $E_z$. The imaginary part of $Z^\parallel$ is invariant
under reversal of the direction of the beam, i.e. $\text{Im}(Z^\parallel)$ is
invariant under the replacement $R(z) \mapsto R(-z)$. Thus, each term
in the integrand in $\text{Im}(Z^\parallel)$ only contains an even number of
derivatives of $R$ and arises from an even power of $\epsilon$ in the
asymptotic series for $E_z$. It follows that the odd order
contributions to $\text{Im}(Z^\parallel)$ vanish.
%
\subsection{Impedance of an off-axis beam}
This section focusses on longitudinal impedance due to the off-axis harmonic
line charge density
\begin{equation}
\label{offset_line_source}
\varrho_{\omega,r_0,\theta_0}(r,\theta,u) =
\frac{I_\omega}{c} e^{i\omega u}\delta\left(x-r_0\cos \theta_0\right)\delta\left(y-r_0\sin\theta_0\right)
\end{equation}
where $r_0 \neq 0$. The corresponding transverse impedance ${\bf
  Z}_\perp$ may be obtained using the Panofsky-Wenzel
relation~\cite{panofsky56}.
 
Fourier expansion of the source in $\theta$ leads to
\begin{align}
Z^\parallel(\omega) = Z^\parallel_{\text{on-axis}} + Z^\parallel_1 + Z^\parallel_2
+ Z^\parallel_4 + Z^\parallel_6 + \dots
\end{align}
where $Z^\parallel_{\text{on-axis}}$ (see (\ref{monopole_Z}))
is the monopole contribution to $Z^\parallel$ and the
contributions arising from 1st, 2nd, 4th and 6th-order terms in $E_z$
are
\begin{align}
\label{imp2off2}
&Z^\parallel_1 =
\frac{1}{4\pi\varepsilon_0
  c}\sum_{m=1}^{\infty}\frac{2}{m}\bigg(\frac{1}{R_2^{2m}}
-\frac{1}{R_1^{2m}}\bigg)\,r^m r_0^m \cos m(\theta-\theta_0),\\
\label{imp2off2_2}
&Z^\parallel_2 =
\frac{1}{4\pi\varepsilon_0 c^2}\sum_{m=1}^{\infty}\Bigg\{-\frac{4i\omega}{1+m}\int_{-\infty}^\infty{\left(\frac{
    R'}{R^{m}}\right)^2\ud z \Bigg\}} r^m r_0^m \cos m(\theta-\theta_0),\\
&4\pi\varepsilon_0 c Z^\parallel_4= \sum_{m=1}^{\infty} \frac{2i\omega}{c}\frac{r_0^m r^m \cos
 m(\theta-\theta_0)}{m(m+1)(m+2)}\Bigg\{ \intfin{\left(\frac{
 R'^4}{R^{2m}}\frac{2m^2+6m+1}{3}+\frac{
 R''^2}{R^{2m-2}}\right)\ud z}\nn\\
\label{imp2off2_4}
&\qquad
 -\frac{\omega^2}{c^2}\frac{3m^2+8m+6}{m(m+1)^2(m+3)}\intfin{\left(\frac{4m-3}{3}\frac{R'^4}{R^{2m-2}}+\frac{R''^2}{R^{2m-4}}\right)\ud z} \Bigg\},
\end{align}
\begin{align}
\notag
4\pi \varepsilon_0 c Z^\parallel_6 =&
\sum_{m=1}^{\infty}\frac{ i\omega}{c}\frac{r_0^m r^m \cos
 m(\theta-\theta_0)}{10m^2(m+1)^3(m+2)(m+3)}\int_{-\infty}^{\infty}\Bigg\{
5(m^2+8m+6)\frac{R'''^2}{R^{2m-4}}\\
\notag
&+ 5(10m^3+14m^2-9m-18)\frac{R''^3}{R^{2m-3}}
+ 15m(2m^3+20m^2+41m+27)\frac{R''^2
  R'^2}{R^{2m-2}} \\
\notag
&+2m(4m^5+32m^4+99m^3+137m^2+85m+15)\frac{R'^6}{R^{2m}}
+\frac{\omega^2}{c^2 m(m+4)}\Bigg[
  30(5m^2+20m+24)\frac{R'''^2}{R^{2m-6}}\\
\notag
&+20(23m^3+40m^2-58m-156)\frac{R''^3}{R^{2m-5}}
+10(236m^3+693m^2+304m-816)\frac{R'^2
  R''^2}{R^{2m-4}}\\
\notag
&+4(154m^4+375m^3-2m^2-628m+240)\frac{R'^6}{R^{2m-2}}
+\frac{\omega^2}{c^2 m(m+1)^2(m+2)(m+5)}\times\\
\notag
&\Bigg[-5(35m^5 + 351 m^4 + 1428m^3+2836m^2 + 2672m +
  960)\frac{R'''^2}{R^{2m-8}}\\
\notag
&-5(106m^6+756m^5+1639m^4-923m^3-8636m^2-11440m-4800)\frac{R''^3}{R^{2m-7}}\\
\notag
&+5(30m^7-378m^6-4395m^5-14574m^4-13667m^3
+19680m^2+43424m+21120)\frac{R'^2
  R''^2}{R^{2m-6}}\\
\notag
&+(m-1)(60m^7-33m^6-2713m^5-10453m^4-9129m^3\\
\label{imp2off2_6}
&+19912m^2+40000m+19200)\frac{R'^6}{R^{2m-4}}\Bigg]\Bigg]\Bigg\}\ud z
\end{align}
respectively, where $m$ is the multipole index. Since
$\text{Im}(Z^\parallel)$ is invariant under the replacement $R(z)\mapsto R(-z)$,
i.e. invariant under reversal of the direction of the beam,
the odd order terms in the asymptotic expansion of $\text{Im}(Z^\parallel)$ in $\epsilon$ vanish. 

The longitudinal impedance of a dipole source is obtained by setting
$m=1$ in (\ref{imp2off2}, \ref{imp2off2_2}, \ref{imp2off2_4}, \ref{imp2off2_6}):
\begin{align}
\label{imp2off2_dip}
&4\pi\epsilon_0 c Z_1^\parallel =
2\left(\frac{1}{R_2}-\frac{1}{R_1}\right) r_0 r\cos
(\theta-\theta_0),\\
\label{imp2off2_2_dip}
&4\pi\epsilon_0 c Z_2^\parallel = -2i\frac{\omega}{c} r_0 r\cos
(\theta-\theta_0) \int_{-\infty}^\infty \frac{R'^2}{R^2}\ud z,\\
\label{imp2off2_4_dip}
&4\pi\epsilon_0 c Z_4^\parallel =i\frac{\omega}{c} r_0 r\cos
(\theta-\theta_0) \Bigg[ \int_{-\infty}^{\infty}
{\left(\frac{{R'}^4}{R^2}+ \frac{1}{3} {R''}^2\right)\ud z}
-\frac{\omega^2}{c^2}\int_{-\infty}^{\infty}{
\left(\frac{17}{144}{R'}^4+\frac{17}{48}{R''}^2 R^2\right)\ud
z}\Bigg],\\
\notag
&4\pi\epsilon_0 c Z_6^\parallel = i\frac{\omega}{c} r_0 r\cos (\theta-\theta_0) \Bigg[
\int_{-\infty}^\infty {\left( -\frac{31}{40}\frac{{R'}^6}{R^2} -
  \frac{45}{32}{R'}^2 {R''}^2 - \frac{17}{192}{R'''}^2 R^2+
  \frac{1}{64}{R''}^3R\right) \ud z}\\
\notag
&\qquad\qquad + \frac{\omega^2}{c^2} \int_{-\infty}^\infty {\left(
  \frac{139}{1200}{R'}^6 + \frac{139}{160} {R''}^2 {R'}^2 R^2
  +\frac{49}{160} {R'''}^2 R^4 - \frac{151}{240} {R''}^3 R^3\right)\ud
  z }\\
\label{imp2off2_6_dip}
&\qquad\qquad + \frac{\omega^4}{c^4} \int_{-\infty}^\infty
  {\left(-\frac{4141}{34560}{R'''}^2 R^6 + \frac{3883}{11520} {R''}^3
  R^5 + \frac{427}{576} {R''}^2 {R'}^2 R^4\right)\ud z}\Bigg].
\end{align}
Use of the Panofsky-Wenzel relation~\cite{panofsky56} and
(\ref{imp2off2_dip}, \ref{imp2off2_2_dip}, \ref{imp2off2_4_dip},
\ref{imp2off2_6_dip}) leads to the following contributions to the
transverse dipole impedance $Z^\perp(\omega)$:
\begin{align}
&4\pi\epsilon_0 c Z_1^\perp = \frac{2c}{\omega}\left(\frac{1}{R_2}-\frac{1}{R_1}\right), \\
&4\pi\epsilon_0 c Z_2^\perp = -2i \int_{-\infty}^\infty \frac{R'^2}{R^2}\ud z,\\
&4\pi\epsilon_0 c Z_4^\perp =i \Bigg[ \int_{-\infty}^{\infty}
{\left(\frac{{R'}^4}{R^2}+ \frac{1}{3} {R''}^2\right)\ud z}
-\frac{\omega^2}{c^2}\int_{-\infty}^{\infty}{
\left(\frac{17}{144}{R'}^4+\frac{17}{48}{R''}^2 R^2\right)\ud
z}\Bigg], 
\end{align}
\begin{align}
\notag
4\pi\epsilon_0 c Z_6^\perp =& i \Bigg[
\int_{-\infty}^\infty {\left( -\frac{31}{40}\frac{{R'}^6}{R^2} -
  \frac{45}{32}{R'}^2 {R''}^2 - \frac{17}{192}{R'''}^2 R^2+
  \frac{1}{64}{R''}^3R\right) \ud z}\\
\notag
&+ \frac{\omega^2}{c^2} \int_{-\infty}^\infty {\left(
  \frac{139}{1200}{R'}^6 + \frac{139}{160} {R''}^2 {R'}^2 R^2
  +\frac{49}{160} {R'''}^2 R^4 - \frac{151}{240} {R''}^3 R^3\right)\ud
  z }\\
&+ \frac{\omega^4}{c^4} \int_{-\infty}^\infty {\left(-\frac{4141}{34560}{R'''}^2 R^6 + \frac{3883}{11520} {R''}^3 R^5 + \frac{427}{576} {R''}^2 {R'}^2 R^4\right)\ud z}\Bigg]
\end{align}
where
\begin{equation}
\label{Z_perp}
Z^\perp(\omega) = Z^\perp_1 + Z^\perp_2 + Z^\perp_4 + Z^\perp_6 + \dots.
\end{equation}
The imaginary part of the transverse impedance at zero frequency
$\lim_{\omega\rightarrow 0}\text{Im}[Z^\perp(\omega)]$ was previously
obtained in~\cite{podobedov06}. As far as we are aware, the above expressions
for transverse impedance \emph{as a function of frequency} are new.
\begin{figure}
\centering
\scalebox{0.75}{\includegraphics[width=\textwidth]{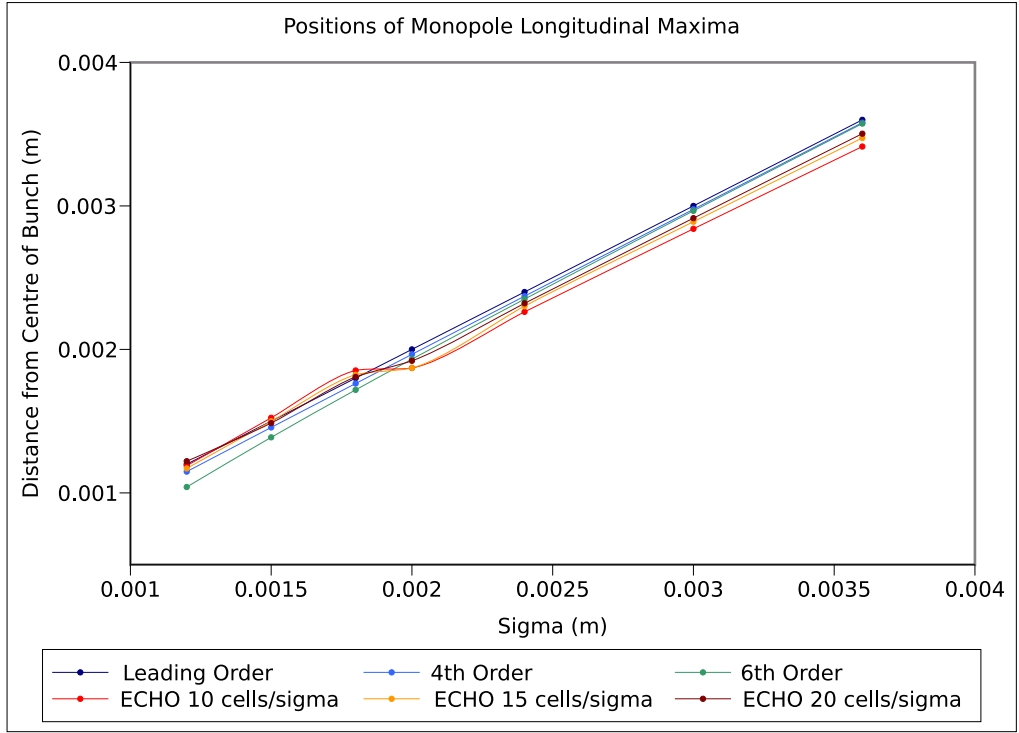}}
\caption{The position of the maximum in $u$ of $W^\parallel$ versus effective bunch length
  $\sigma$ for a monopole source propagating through the exponential geometry (\ref{exp_profile}).}
\end{figure}
\begin{figure}
\centering
\scalebox{0.75}{\includegraphics[width=\textwidth]{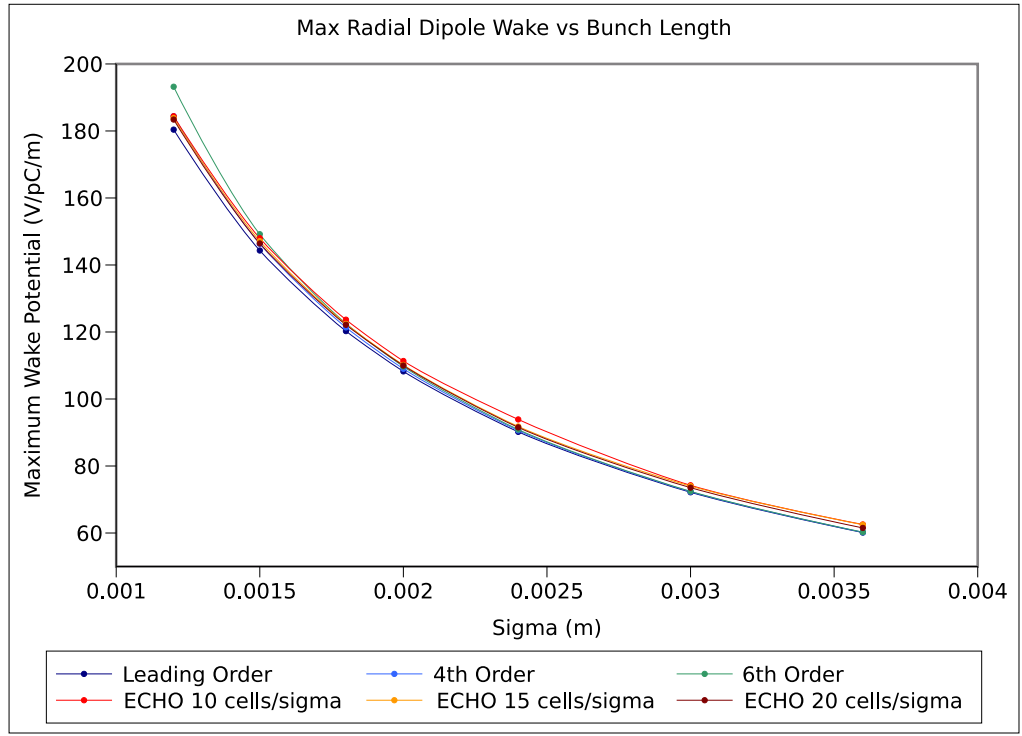}}
\caption{The maximum in $u$ of the radial component of ${\bf W}_\perp$
  versus effective bunch length $\sigma$ for a dipole source
  propagating through the exponential geometry (\ref{exp_profile}).}
\end{figure}
%
\begin{figure}
\centering
\scalebox{0.75}{\includegraphics[width=\textwidth]{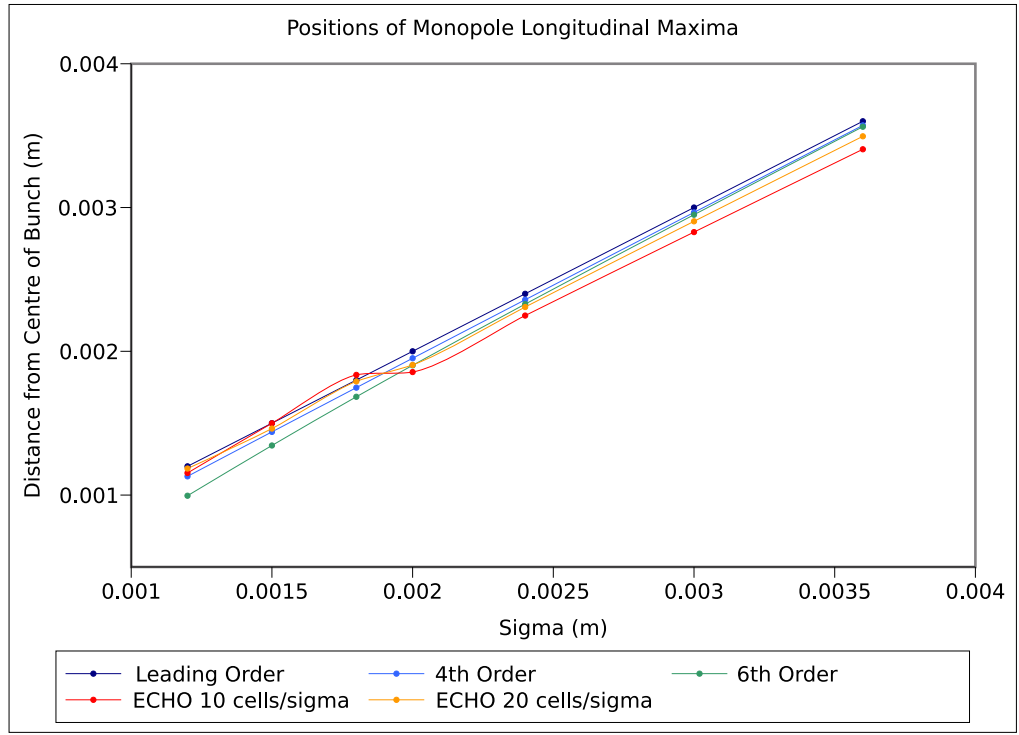}}
\caption{The position of the maximum in $u$ of $W^\parallel$ versus effective bunch length
  $\sigma$ for a monopole source propagating through the hyperbolic
  secant geometry (\ref{sech_profile}).}
\end{figure}
\begin{figure}
\centering
\scalebox{0.75}{\includegraphics[width=\textwidth]{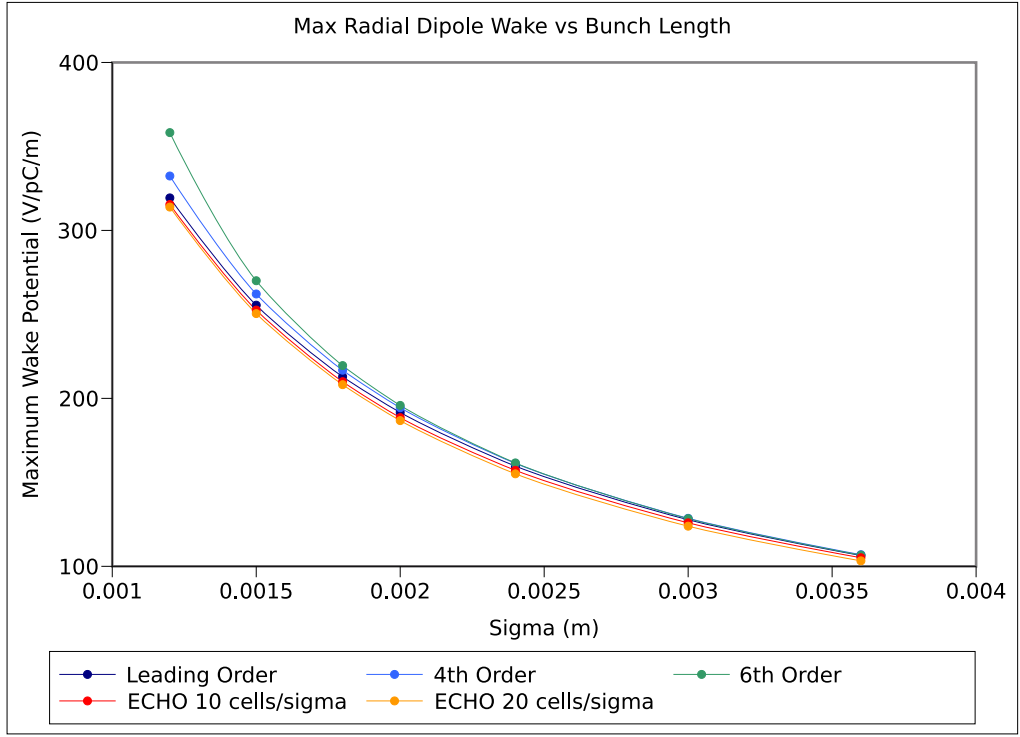}}
\caption{The maximum in $u$ of the radial component of ${\bf W}_\perp$
  versus effective bunch length $\sigma$ for a dipole source
  propagating through the hyperbolic secant geometry
  (\ref{sech_profile}).}
\end{figure}
\section{Derivation of results}
The remainder of the paper focusses on the method used to derive the
expressions presented above.
\subsection{Decomposition of Maxwell's equations using auxiliary potentials}
\label{section:decomp_aux_potentials}
\subsubsection{2+2 split of Maxwell's equations}
The approach we use is based on a 2+2 split of Maxwell's equations
adapted to the waveguide. We employ
cylindrical polar co-ordinates $(t,r,\theta,z)$ on Minkowski spacetime
where $r=0$ is the waveguide's axis, $\theta$ is the azimuthal angle around
the waveguide's axis and $z$ is a co-ordinate parallel to
the waveguide's axis. We employ exterior differential calculus to
analyse Maxwell's equations (see, for
example,~\cite{benn_tucker87, burton03}) as it affords a concise language for
exploiting different coordinate systems and for developing our new
auxiliary potential method discussed in the following. 

Let $({\cal M},g)$ be Minkowski spacetime where $g$ is the metric
\begin{equation}
g =-c^2 \ud t \otimes \ud t + \ud z \otimes \ud z+ \ud r \otimes \ud r +
r^2\ud\theta \otimes\ud\theta
\end{equation}
in $(t,r,\theta,z)$ co-ordinates. We choose the volume $4$-form $\star 1$ as
\begin{equation}
\star 1 = c\ud t\wedge \ud z\wedge \hp 1
\end{equation}
where the $2$-form $\hp 1$,
\begin{equation}
\hp 1 = r^2 \ud r\wedge \ud\theta,
\end{equation}
and the metric $g_{\perp}$,
\begin{equation}
g_{\perp} =  \ud r \otimes \ud r + r^2\ud\theta \otimes\ud\theta, 
\end{equation}
are adapted to constant $(t,z)$ surfaces inside the waveguide. The
waveguide surface in ${\cal M}$ is the $3$-dimensional hypersurface
$\{r = R(z)\,|\,-\infty<z<\infty\}$ where $R(z)$ is the
radius of the cross-section at $z$ with unit normal $\partial_z$.

The source moves in the positive $z$-direction near the speed of
light in the laboratory frame, and its $4$-velocity field $V$ is
well-approximated by the null vector :
\begin{equation}\label{vfield}
V = \frac{1}{c}\partial_t + \partial_z.
\end{equation}
The metric isomorphism between the spaces of vectors and co-vectors on spacetime
is denoted as follows: the co-vector field $\widetilde{X}$ associated with
the vector field $X$ is defined such that $\widetilde{X}(Y) =
g(X,Y)$ for all vector fields $Y$ and the vector field $\widetilde{\alpha}$
associated with the co-vector field $\alpha$ is defined such that
$\beta(\widetilde{\alpha}) = g^{-1}(\beta,\alpha)$ for all co-vector fields
$\beta$. Hence, the $1$-form $\widetilde{V}$ associated with $V$ is
\begin{equation}
\widetilde{V} = \ud z - c\ud t.
\end{equation}

It is expedient to transform to a new co-ordinate chart:
\begin{equation}\label{uzeta}
u = z-ct, \qquad \zeta = z
\end{equation}
with
\begin{align}
&g = \ud\zeta \otimes \ud u + \ud u \otimes \ud \zeta - \ud u \otimes
\ud u + g_{\perp},\label{seminull}\\
&\star 1 = \ud \zeta \wedge \ud u \wedge \hp 1.
\end{align}
The coordinates (\ref{uzeta}) play a useful role because the source is
naturally expressed in terms of $u$ and the boundary of the waveguide
is described using $\zeta$. It follows that
\begin{align}
&\vt = \ud u,\label{vtu}\\
&\star\vt = \ud u \wedge \hp 1.
\end{align}
Here, the Hodge map $\star$ on $\cal{M}$ satisfies
\begin{equation}
 \star(\nu\wedge \widetilde{X}) = \ui_X\star\nu
\end{equation}
where $\nu$ is an arbitrary form and $X$ is an arbitrary vector field
and $\ui_X$ is the interior contraction with $X$. The \emph{transverse Hodge map} $\hp$ is defined on the transverse, cross-sectional subspace to satisfy
\begin{equation}
 \hp(\mu_\perp\wedge \widetilde{Y}) = \ui_Y\hp\mu_{\perp}
\end{equation}
where $\mu_\perp$ is a form on the domain spanned by $\{\ud r, \ud\theta\}$ and $Y$ is a vector field in the span of $\{\partial_r, \partial_\theta\}$.

The Maxwell equations on space-time are
\begin{align}
&\ud F = 0\label{max1}\\
&\varepsilon_0 \ud \star F = -\varrho \star \vt\label{max2}
\end{align}
where $\varrho\widetilde{V}$ is the electric $4$-current of the beam and $\ud$ is
the exterior derivative on forms. Charge conservation
\begin{equation}
\ud(\varrho \star \vt) = 0
\end{equation}
yields 
\begin{equation}\label{integrability}
\partial_\zeta \varrho =0.
\end{equation}

Without loss of generality, the electromagnetic $2$-form $F$ is decomposed as
\begin{equation}\label{F_0}
F = \Phi \ud\zeta \wedge\ud u + \ud u \wedge \alpha +
\ud\zeta\wedge\bp + \Psi \hp 1
\end{equation}
where $\alpha$ and $\bp$ lie in the subspace of
$1$-forms spanned by $\{\ud r,\,\ud\theta\}$ (transverse
subspace). The component $\Phi$ of $F$ is the longitudinal component
of the electric field in $\left(r,\theta,u,\zeta\right)$ coordinates:
\begin{equation}
E_z\left(r,\theta,z=\zeta,t=\frac{1}{c}(\zeta-u)\right) = \Phi(r,\theta,u,\zeta).
\end{equation}
(See, for example,~\cite{burton07} for a detailed discussion of the relationship
between $F$ and electric and magnetic fields in vector notation). 
It follows that the Maxwell equations (\ref{max1}) and (\ref{max2}) can be decomposed as
\begin{align}
&\dpe \Phi + \partial_\zeta
\alpha - \partial_u \bp = 0,\label{maxu1}\\
&\partial_\zeta\Psi \hp 1 - \dpe \bp = 0,\label{maxu2}\\
&\partial_u \Psi \hp 1 - \dpe \alpha = 0,\label{maxu3}\\
&\partial_\zeta \hp (\alpha + \bp
)+\partial_u\hp\bp - \dpe \Psi = 0,\label{maxu4}\\
&\dpe \hp(\alpha + \bp)-\partial_u\Phi\hp
1 = -\rho\rtu\hp1,\label{maxu5}\\
&\partial_\zeta\Phi \hp 1 + \dpe \hp
\bp = 0\label{maxu6}
\end{align}
where $\dpe$ is the exterior derivative acting in the transverse
subspace, and $\rho(r,\theta,u)\equiv\frac{1}{\varepsilon_0}\varrho(r,\theta,u)$, the functional dependence of the charge density being written explicitly to emphasize its independence of $\zeta$ due to (\ref{integrability}). 
%
\subsubsection{Auxiliary potentials}
\label{section:auxiliary_potentials}
In this section we introduce a set of six 0-forms (\emph{auxiliary potentials}).  This anticipates the imposition of the perturbation scheme in Section \ref{gradualsec}: they will be shown to satisfy tractable second order PDEs in
gradually tapering waveguides, whereas working directly with the conventional electromagnetic potentials leads to third order PDEs. Appendix \ref{juste}
contains a brief discussion of this issue.

After rewriting the Maxwell equations in terms of auxiliary
potentials, and adopting perfectly conducting boundary conditions, we will
show that asymptotic expansions of the auxiliary potentials yield 2-dimensional
Poisson and Laplace equations at each order in the expansion parameter
$\epsilon$.

In Appendix \ref{decompose}, we demonstrate that the transverse 1-forms $\alpha$ and
$\bp$ can be replaced by the $0$-form pairs $\{A,a\}$ and $\{B,b\}$, viz.
\begin{align}
&\alpha = \dpe A  +  \hp\dpe a,\label{alpdec}\\
&\bp  = \dpe B  + \hp\dpe b\label{betdec}
\end{align}
where $A$ and $B$ obey Dirichlet boundary conditions on the waveguide
boundary, i.e. $A=B=0$ at $r=R(\zeta)$.



It follows that (since $\ud^2=0$) the  Maxwell equations (\ref{maxu1}-\ref{maxu6}) partially decouple to give
\begin{align}
&\dpe \left(\Phi + \partial_\zeta
A - \partial_u B\right) + \hp\dpe\left(\partial_\zeta
a - \partial_u b\right) = 0,\label{maxs1}\\
&\partial_\zeta\Psi  +\cp\dpe b = 0,\label{maxs2}\\
&\partial_u \Psi +\cp\dpe a = 0,\label{maxs3}\\
&\dpe\left(\partial_\zeta  (A + B) +\partial_u B\right)
+ \hp\dpe\left(\Psi + \partial_\zeta(a+b)+\partial_u b\right) = 0,\label{maxs4}\\
&\cp\dpe(A+B)+\partial_u\Phi = \rho\rtu,\label{maxs5}\\
&\partial_\zeta\Phi - \cp\dpe B = 0,\label{maxs6}
\end{align}
where
\begin{equation}
\cp\mu_\perp \equiv (-1)^p \hp^{-1}\dpe\hp\mu_\perp \end{equation} is the
transverse co-derivative of a transverse (i.e. in the span of $\{\ud
r, \ud\theta\}$) $p$-form $\mu_\perp$ and $\hp^{-1}\hp = \hp\hp^{-1} = 1$.

Motivated by the appearance of equations (\ref{maxs1}) and
(\ref{maxs4}), introduce $0$-forms $\{\hr^\Phi$, $\hr^\varphi$,
$\hr^B$, $\hr^b$, $\hW$, $\hX\}$ that satisfy
\begin{align}
&\partial_u\hr^\Phi = \Phi + \partial_\zeta A - \partial_u B,\label{phidef}\\
&\partial_u \hr^\varphi =\partial_\zeta a - \partial_u b,\label{hrvar}\\
&\partial_u \hr^B = \partial_\zeta (A+B) + \partial_u B,\label{hrb}\\
&\hr^b=\Psi + \partial_\zeta(a+b) + \partial_u b,\label{hrbsm}\\
&\partial_u\hW = A+ B,\label{wab}\\
&\partial_u \hX = a.\label{ainx}
\end{align}
Equations (\ref{phidef}-\ref{ainx}) do not specify $\{\hr^\Phi,
\hr^\varphi, \hr^B, \hW, \hX\}$ uniquely, as they all appear under a
$u$-derivative. Additional ``gauge fixing'' conditions will be supplied in the following.

Substituting (\ref{wab}) into (\ref{hrb}) and integrating with respect
to $u$ gives
\begin{equation}
B=\hr^B - \partial_\zeta\hW + \lambda\rtz\label{Bdef}
\end{equation}
where $\lambda\rtz$ is an arbitrary $0$-form of integration and,
hence, from (\ref{phidef}, \ref{wab}, \ref{Bdef}),
\begin{align}
& A = \partial_u \hW + \partial_\zeta \hW - \hr^B - \lambda\rtz\label{Adef},\\
&\Phi = \partial_u \hr^\Phi + \partial_u \hr^B + \partial_\zeta \hr^B
  + \partial_\zeta \lambda\rtz
 - 2 \partial^2_{u\zeta}\hW - \partial^2_{\zeta\zeta}\hW.\label{phiw}
\end{align}
Substituting (\ref{ainx}) into (\ref{hrvar}) and integrating with
respect to $u$ gives
\begin{equation}
b=\partial_\zeta \hX - \hr^\varphi-\gamma\rtz  \label{binx}
\end{equation}
where $\gamma\rtz$ is a $0$-form of integration. Substituting
(\ref{ainx}) and (\ref{binx}) into (\ref{hrbsm}) gives
\begin{align}\label{pinx}
\Psi = -2\partial^2_{\zeta u}\hX - \partial^2_{\zeta\zeta} \hX +   \partial_\zeta \hr^\varphi + \partial_u\hr^\varphi + \partial_\zeta\gamma\rtz + \hr^b.
\end{align}
We can simplify the above by introducing $0$-forms $\Lambda\rtz$ and
$\Gamma\rtz$,
\begin{align}
&\partial_\zeta \Lambda\rtz=\lambda\rtz,\\
&\partial_\zeta \Gamma\rtz=\gamma\rtz
\end{align}
and $0$-forms $W$, $X$,
\begin{align}
&W = \hW - \Lambda\rtz,\label{whwl}\\
&X = \hX - \Gamma\rtz\label{xhxg}.
\end{align}
Substituting the above into (\ref{ainx}-\ref{pinx}) gives
\begin{align}
&B=\hr^B - \partial_\zeta W,\label{beqf}\\
&A = \partial_u W  + \partial_\zeta W- \hr^B,\label{aeqf}\\
&\Phi = \partial_u \hr^\Phi + \partial_u \hr^B + \partial_\zeta \hr^B -
2 \partial^2_{u\zeta} W - \partial^2_{\zeta\zeta}W,\label{phieqf}\\
&a = \partial_u X,\label{saeqf}\\
&b=\partial_\zeta X - \hr^\varphi,\label{sbeqf}\\
&\Psi = -2\partial^2_{\zeta u}X- \partial^2_{\zeta\zeta}X +
\partial_\zeta \hr^\varphi + \partial_u\hr^\varphi +
\hr^b.\label{psieqf}
\end{align}
Having derived a mapping from the auxiliary potentials
$\{\hr^B, W, \hr^\Phi, X,\hr^\varphi\}$ to $\{B, A, \Phi, a, b,
\Psi\}$ we now rewrite equations
(\ref{maxs1}-\ref{maxs6}) in terms of $\{\hr^B, W, \hr^\Phi,
X,\hr^\varphi\}$.
%
Equations (\ref{maxs1}) and (\ref{maxs4}) become
\begin{align}
&\partial_u\left(\dpe\hr^\Phi + \hp\dpe\hr^\varphi\right)=0,\label{mh1}\\
&\partial_u \dpe\hr^B + \hp\dpe\hr^b=0 \label{mh4}.
\end{align}
Equation (\ref{mh1}) is equivalent to
\begin{equation}\label{kapint}
\dpe\hr^\varphi = \hp\dpe\hr^\Phi +\kappa\rtz
\end{equation}
where $\kappa\rtz$ is an integration $1$-form. Decomposing $\kappa$ as
(see Appendix~\ref{decompose})
\[ \kappa\rtz = \hp\dpe\kappa_1\rtz - \dpe\kappa_2\rtz\]
and using (\ref{kapint}) yields
\begin{equation}\label{kapint2}
\dpe\left(\hr^\varphi + \kappa_2\rtz \right)= \hp\dpe\left( \hr^\Phi + \kappa_1\rtz\right).
\end{equation}

It can be seen from (\ref{beqf}-\ref{psieqf}) that the
$0$-forms $A$, $a$, $B$, $b$, $\Phi$, $\Psi$ (and hence our
electromagnetic 2-form $F$) are invariant under the ``gauge''
transformations~\footnote{Unrelated to the usual electromagnetic gauge transformations.}
\begin{align}
&\hr^B \rightarrow \hr^B + \partial_\zeta Q\rtz,\\
&W \rightarrow W + Q\rtz + \hat w(r,\theta), \label{gaugew}\\
&\hr^\Phi\rightarrow \hr^\Phi+n\rtz,\\
&\hr^\varphi \rightarrow \hr^\varphi + \partial_\zeta M\rtz,\\
&X \rightarrow X + M\rtz + \hat x(r,\theta)\label{gaugex}
\end{align}
where $Q$, $\hat w$, $n$, $M$ and $\hat x$ are arbitrary functions of the indicated variables.  In what follows, these ``gauge functions'' will be chosen to reduce the Maxwell
equations to a form amenable to the gradual taper approximation.

Without imposing any restrictions on the electromagnetic 2-form $F$, we choose $\{M,n\}$ so that
$\{\kappa_1,\kappa_2\}$ satisfy the ``gauge'' conditions
\begin{align}
&\partial_\zeta M\rtz = -\kappa_2\rtz\\
&n \rtz = -\kappa_1\rtz.
\end{align}
Equation (\ref{kapint2}) becomes
\begin{equation}\label{phiha1}
\dpe\hr^\varphi = \hp\dpe\hr^\Phi. 
\end{equation}
Applying the transverse exterior derivative $\dpe$ and co-derivative
$\cp$ to (\ref{phiha1}) gives
\begin{equation}
\label{H_phi_harmonic}
\cp\dpe\hr^\Phi=\cp\dpe\hr^\varphi=0.
\end{equation}
Similarly, equation (\ref{mh4}) implies
\begin{equation}
\cp\dpe\hr^B=\sigma\rtz
\end{equation}
where $\sigma\rtz$ is a $0$-form of integration and so, by choosing
$Q\rtz$ such that $\partial_\zeta\cp\dpe Q\rtz=\sigma\rtz$, we find
\begin{equation}\label{harmb}
\cp\dpe\hr^B=0.\end{equation}
Thus, $\{\hr^\Phi$, $\hr^\varphi$, $\hr^B$, $\hr^b\}$ may be chosen to be harmonic with
respect to the 2-dimensional transverse Laplacian. 
Now turning to (\ref{maxs2}, \ref{maxs3}), we note that 
the Maxwell equations (\ref{maxs3}) and (\ref{saeqf}) give
\begin{equation}
\partial_u\left(\cp\dpe X + \Psi\right)=0
\end{equation}
so that
\begin{equation}\label{intm2}
\cp\dpe X + \Psi = \nu\rtz
\end{equation} where $\nu\rtz$ is a $0$-form of
integration.  As $\hr^\varphi$ is harmonic (see
(\ref{H_phi_harmonic})), the Maxwell equations (\ref{maxs2}) and
(\ref{sbeqf}) give
\begin{equation}
\partial_\zeta\left(\cp\dpe X + \Psi\right)=0
\end{equation}
and it follows
\begin{equation}\label{intm3}
\cp\dpe X + \Psi = \mu\rtu
\end{equation}
where $\mu\rtu$ is an integration $0$-form. The left hand sides of
(\ref{intm2}) and (\ref{intm3}) are equal, so
\begin{equation}
\nu\rtz = \mu\rtu.
\end{equation}
Hence, the integration $0$-forms $\mu\rtu$ and $\nu\rtz$ are constant
with respect to $u,\zeta$. Eliminating $\Psi$ from (\ref{intm2}) using
(\ref{psieqf})
yields
\begin{equation}\label{xwavn}
\cp\dpe X -2\partial^2_{\zeta u}X- \partial^2_{\zeta\zeta}X +
\partial_\zeta \hr^\varphi + \partial_u\hr^\varphi +  \hr^b=
\hat{\mu}(r,\theta) 
\end{equation}
where $\hat\mu(r,\theta) = \mu\rtu = \nu\rtz$.
The function $\hat x(r,\theta)$ has not yet been used in the
transformation (\ref{gaugex}).  We can eliminate $\hat\mu(r,\theta)$ without
affecting the fields by setting 
\begin{equation}
\cp\dpe \hat x(r,\theta)= \hat \mu(r,\theta).
\end{equation}
Thus, we conclude that (\ref{maxs2}) and (\ref{maxs3}) may be
written as the single equation
\begin{equation}\label{xwave}
\boxed{\cp\dpe X -2\partial^2_{\zeta u}X- \partial^2_{\zeta\zeta}X +   \partial_\zeta \hr^\varphi + \partial_u\hr^\varphi +  \hr^b= 0.}
\end{equation}
We now consider the final two Maxwell equations (\ref{maxs5}, \ref{maxs6}).
Using (\ref{beqf}, \ref{aeqf}), the Maxwell equation
(\ref{maxs5}) may be written as
\begin{equation}\label{m2st}
\partial_u \left(\cp\dpe W + \Phi\right) = \rho\rtu.
\end{equation}
Introduce $P\rtu$ where
\begin{equation}\label{prho}
\partial_u P\rtu = \rho\rtu.
\end{equation}
Equation (\ref{m2st}) can thus be integrated to give
\begin{equation}\label{m561}
\cp\dpe W + \Phi =P\rtu + \varsigma\rtz
\end{equation}
where $\varsigma\rtz$ is a $0$-form of integration. Now, using
(\ref{beqf}), and noting that $\hr^\varphi$ is harmonic, (\ref{maxs6})
gives
\begin{equation}
\partial_\zeta\left(\cp\dpe W + \Phi\right) = 0
\end{equation}
which implies
\begin{equation}\label{m562}
\cp\dpe W + \Phi=\chi\rtu
\end{equation}
where $\chi\rtu$ is a $0$-form of integration.
Equating the right hand sides of (\ref{m561}) and (\ref{m562}) gives
\begin{equation}
\chi\rtu = P\rtu + \varsigma\rtz.
\end{equation}
Thus, $\varsigma\rtz$ is constant with respect to $\zeta$
and eliminating $\Phi$ from (\ref{m561}) using (\ref{phieqf}) yields
\begin{align}
\label{fww4a}
\cp\dpe W - 2 \partial^2_{u\zeta} W - \partial^2_{\zeta\zeta} W +
\partial_u\hr^\Phi + \partial_u \hr^B + \partial_\zeta \hr^B  =  P\rtu+ \hat\varsigma(r,\theta)
\end{align}
where $\hat\varsigma(r,\theta) = \varsigma\rtz$. Using the transformation
(\ref{gaugew}) we may set
\[\cp\dpe \hat w(r,\theta)  = \hat\varsigma(r,\theta)\]
and conclude that, without loss of generality in $F$, one may choose a
``gauge'' in which
\begin{equation}
\boxed{
\label{fww4}
\cp\dpe W - 2 \partial^2_{u\zeta} W - \partial^2_{\zeta\zeta} W
+   \partial_u\hr^\Phi + \partial_u \hr^B
+ \partial_\zeta \hr^B  =  P\rtu
}
\end{equation}
is satisfied.
\subsubsection{Summary}
We
seek appropriately bounded solutions to the following system of equations for our auxiliary potentials $W$, $X$, $\hr^B$, $\hr^b$, $\hr^\Phi$ and $\hr^\varphi$:
\begin{align}
&\cp\dpe\hr^B =0\label{fg1},\\
&\dpe \hr^b = \hp\dpe\left(\partial_u \hr^B\right),\label{fg2}\\
&\dpe\hr^\varphi = \hp\dpe\hr^\Phi,\label{fg3}\\
&\cp\dpe W - 2 \partial^2_{u\zeta} W - \partial^2_{\zeta\zeta} W
+ \partial_u\hr^\Phi + \partial_u \hr^B + \partial_\zeta \hr^B  =  P\rtu,\label{fg4}\\
&\cp\dpe X -2\partial^2_{u\zeta} X - \partial^2_{\zeta\zeta} X + \partial_\zeta \hr^\varphi + \partial_u\hr^\varphi+\hr^b =0.\label{fg5}
\end{align}
Notice that (\ref{fg2}) and (\ref{fg3}) imply that $\cp\dpe\hr^b=\cp\dpe\hr^\Phi=\cp\dpe\hr^\varphi=0$.  The source term $P\rtu$ in equation (\ref{fg4}) satisfies equation (\ref{prho}):
\[
\partial_u P\rtu=\rho\rtu
\]
and the electromagnetic 2-form $F$ is
\begin{align}
F = \Phi \ud\zeta \wedge\ud u &+ \ud u \wedge\left(\dpe A + \hp\dpe
a\right) + \ud\zeta\wedge\left(\dpe B + \hp\dpe b\right) + \Psi \hp 1
\label{F_in_terms_of_A}
\end{align}
where 
\begin{align} 
&A =\partial_u W + \partial_\zeta W - \hr^B, \label{Afin}\\
&B =  \hr^B - \partial_\zeta, W\label{Bfin}\\
&\Phi =  \partial_u\hr^\Phi + \partial_u \hr^B + \partial_\zeta \hr^B
  - 2 \partial^2_{u\zeta} W - \partial^2_{\zeta\zeta}
  W,\label{phifin}\\
&a = \partial_u X, \label{fina}\\
&b=\partial_\zeta  X - \hr^\varphi, \label{finb}\\
&\Psi = -\left(2\partial^2_{u\zeta}X + \partial^2_{\zeta\zeta}X\right)  + \partial_\zeta \hr^\varphi +\partial_u \hr^\phi+\hr^b.\label{finpsi}
\end{align}
Equations (\ref{fg4}) and (\ref{fg5}) also give alternative expressions for $\Phi$ and $\Psi$:
\begin{align}
&\Phi = P\rtu - \cp\dpe W,\label{phialt}\\
&\Psi = -\cp\dpe X\label{psialt}
\end{align}
and $F$ may be written
\begin{align}
&F = \left(\partial_u\hr^\Phi + \partial_u \hr^B + \partial_\zeta \hr^B - 2 \partial^2_{u\zeta} W - \partial^2_{\zeta\zeta} W\right)\ud\zeta\wedge\ud u\nn\\
&+ \ud u \wedge\left[\dpe\left(\partial_u W + \partial_\zeta W -
  \hr^B\right) + \partial_u\hp\dpe X\right]\nn\\
&+ \ud \zeta \wedge\left[\dpe\left(\hr^B - \partial_\zeta W\right) +
    \hp\dpe\left(\partial_\zeta  X - \hr^\varphi \right)\right]\nn\\
&+   \left(\partial_\zeta \hr^\varphi +\partial_u \hr^\phi+\hr^b -
2\partial^2_{u\zeta}X - \partial^2_{\zeta\zeta}X\right)\hp
1\label{Fwx1}
\end{align}
At first glance, it may seem that (\ref{fg1}-\ref{fg5}) is no more
amenable to analysis than the original Maxwell equations.
However, we show in the following that these
equations reduce to 2nd order PDEs that are straightforward to solve
when the auxiliary potentials are slowly varying with
$\zeta$.
%
\subsection{Boundary conditions on the auxiliary potentials}
The boundary of the waveguide is the surface
\begin{equation}\label{surf}
r - R(\zeta)=0.
\end{equation}
Given a function $\phi(r,\theta,\zeta,u)$ such that $\phi(R(\zeta),\theta, \zeta, u)=0$, it follows immediately that
\begin{align}
&\frac{\partial}{\partial \theta}\phi(R(\zeta),\theta, \zeta, u)=0,\label{dthk}\\
&\frac{\partial}{\partial u}\phi(R(\zeta),\theta, \zeta, u)=0,\label{duk}\\
&\frac{\ud}{\ud \zeta}\phi(R(\zeta),\theta, \zeta, u)=\left(\frac{\partial}{\partial \zeta} +
  R'(\zeta)\frac{\partial}{\partial r}\right)\phi(R(\zeta),\theta, \zeta, u)=0.\label{E3}
\end{align}
where $R'(\zeta) = dR/d\zeta$.

Orthogonal decomposition (see Appendix \ref{decompose}) of the 1-forms
$\alpha$ and $\beta$
employs potentials $A$ and $B$ that satisfy Dirichlet boundary
conditions on $r=R(\zeta)$. Hence, on the boundary (\ref{Afin}) and (\ref{Bfin}) give
\begin{eqnarray}
\boxed{\hr^B-\partial_\zeta W=0|_{r=R(\zeta)},}\label{hrbbound}\\
\partial_u W=0|_{r=R(\zeta)}.\label{wbound}
\end{eqnarray}
Equation (\ref{hrbbound}) serves as a boundary condition on $\hr^B$,
while (\ref{wbound}) is satisfied by imposing Dirichlet boundary
conditions on $W$:
\begin{equation}\label{wdir}
\boxed{W=0|_{r=R(\zeta)}.}
\end{equation}

The usual conditions on the electric and magnetic fields at a perfectly
conducting surface $\mathcal{S} = 0$ in spacetime may be written
\begin{equation}\label{bound0}
\ud\mathcal S\wedge F=0|_{{\mathcal S} =0}.
\end{equation}
Using $\mathcal S = r - R(\zeta)$ we can decompose (\ref{bound0}) as
\begin{eqnarray}
\Phi \ud r - R'(\zeta)\alpha =0|_{r=R(\zeta)},\label{boun1}\\
\ud r \wedge\bp + \Psi R'(\zeta)\hp 1 =0|_{r=R(\zeta)}\label{boun2}.
\end{eqnarray}
With $\alpha$ given by (\ref{alpdec}), the $\ud\theta$ component of (\ref{boun1}) is
\begin{equation}
\partial_\theta A + r\partial_r a=0|_{r=R(\zeta)}.
\end{equation}
Since $A$ satisfies Dirichlet boundary conditions at $r=R(\zeta)$
it follows
\begin{equation}\label{aneu}
r\partial_r a=0\qquad \Rightarrow \partial^2_{ru}X=0|_{r=R(\zeta)}
\end{equation}
and this can be satisfied by imposing the boundary condition~\footnote{This corresponds to Neumann boundary conditions \emph{on the
    2 dimensional cross-sectional disc} of radius $R(\zeta)$ where
  $\zeta$ is treated as a parameter.  This anticipates the imposition
  of the perturbation scheme in subsequent sections, where $X$ is
  obtained from the 2D Poisson equation in the disc. This is not the
  same as Neumann boundary conditions on the entire waveguide, as
  the vector field normal to the surface is given by
  $\frac{\partial}{\partial \zeta} + R'(\zeta)\frac{\partial}{\partial
    r}$} on $X$:
\begin{equation}\label{xneum}
\boxed{\partial_r X=0|_{r=R(\zeta)}.}
\end{equation}
The $\ud r$ component of (\ref{boun1}) gives
\begin{equation}\label{drbound}
\Phi + R'(\zeta)\left(\frac{1}{r}\partial_\theta a - \partial_r A\right)=0|_{r=R(\zeta)}
\end{equation}
and using (\ref{phidef}, \ref{fina}) to eliminate $\Phi$ and $a$,
(\ref{drbound}) can be rewritten
\begin{equation}
\label{hr_Phi_bound_cond}
\partial_u\hr^\Phi + R'(\zeta)\frac{1}{r}\partial^2_{\theta u}X=0|_{r=R(\zeta)}
\end{equation}
where 
\begin{equation}
\partial_u B=0|_{r=R(\zeta)}, \qquad \partial_\zeta A + R'(\zeta)
\partial_r A=0|_{r=R(\zeta)}
\end{equation}
have been used, which follow from (\ref{duk}, \ref{E3}) and the Dirichlet
conditions $A=B=0|_{r=R(\zeta)}$.

Equation (\ref{hr_Phi_bound_cond}) can be satisfied by imposing the
following boundary condition on $\hr^{\Phi}$:
\begin{equation}\label{hrphibound}
\boxed{\hr^\Phi =- R'(\zeta)\frac{1}{r}\partial_{\theta}X|_{r=R(\zeta)}.}
\end{equation}
We will now argue that (\ref{boun2}) is automatically satisfied without further
conditions on $\{W, X, \hr^B, \hr^\Phi\}$. Since $B=0|_{r=R(\zeta)}$, it
follows $\partial_\theta B=0|_{r=R(\zeta)}$ and (\ref{boun2}) becomes 
\begin{equation}\label{bpsibound}
\partial_r b + R'(\zeta)\Psi=0|_{r=R(\zeta)}.
\end{equation}
Using (\ref{finb}) and (\ref{psialt}) to eliminate $b$ and $\Psi$, and noting
\[ -\cp\dpe X = \frac{1}{r}\partial_r\left(r\partial_r
X\right)+\frac{1}{r^2}\partial^2_{\theta\theta}X \]
equation (\ref{bpsibound}) becomes
\begin{equation}\label{psibou2}
\partial^2_{r\zeta}X-\partial_r\hr^\varphi + R'(\zeta)\left\{\frac{1}{r}\partial_r\left(r\partial_r X\right)+\frac{1}{r^2}\partial^2_{\theta\theta} X\right\}=0|_{r=R(\zeta)}.
\end{equation}
However, equation (\ref{fg3}) implies
\begin{equation}
\partial_r\hr^\varphi = -\frac{1}{r}\partial_\theta \hr^\Phi
\end{equation}
and hence, from (\ref{hrphibound}), 
\begin{equation}\label{finb1}
\partial_r\hr^\varphi = \frac{R'(\zeta)}{r^2}\partial^2_{\theta\theta}X|_{r=R(\zeta)}.
\end{equation}
Furthermore, using (\ref{xneum}, \ref{E3}) with $\phi=r\partial_r X$ gives
\begin{equation}\label{finb2}
\partial^2_{r\zeta}X + R'(\zeta)
\frac{1}{r}\partial_r\left(r\partial_r X\right)=0|_{r=R(\zeta)}
\end{equation}
hence (\ref{psibou2}) is trivially satisfied using (\ref{finb1}, \ref{finb2}). Thus, 
(\ref{boun2}) is satisfied without further conditions on $\{W, X, \hr^B, \hr^\Phi\}$.
  
In summary, the perfectly conducting boundary conditions on the
electromagnetic field at
$r=R(\zeta)$ are satisfied by imposing the following boundary
conditions on $\{W$, $X$, $\hr^B$, $\hr^\Phi\}$:
\begin{equation}\label{finalbound}
\boxed{
\begin{split}
&W=0|_{r=R(\zeta)}, \quad \partial_r X=0|_{r=R(\zeta)},\\
&\hr^B=\partial_\zeta W|_{r=R(\zeta)}, \quad
  \hr^\Phi =- R'(\zeta)\frac{1}{r}\partial_{\theta}X|_{r=R(\zeta)}.
\end{split}
} 
\end{equation}
\subsection{Untapered Waveguide}
Our method for determining $\{W$,$X$,$\hr^B$,$\hr^\Phi\}$ is based on
asymptotic series for fields modelled on those in the untapered
waveguide.

When the waveguide is a regular cylinder with zero taper
($R'(\zeta)=0$), the Dirichlet boundary condition on $W$ yields
$\partial_\zeta W=0|_{r=R}$ (see (\ref{E3})). Thus,
(\ref{finalbound}) immediately gives
\begin{equation}
\hr^B = \hr^\Phi=0|_{r=R}.
\end{equation}
The auxiliary potentials $\{\hr^B,\hr^\Phi\}$ satisfy the transverse Laplace equation and
$\{\hr^B,\hr^\Phi\}$ vanish at $r=R(\zeta)$. Thus, it follows that
\begin{align}
\notag
\int_\doma{\dpe \hr^B \wedge \hp\dpe\hr^B} &= \int_\domb{\hr^B
  \hp\dpe\hr^B}\\
\notag
&\,- \int_{\doma}{\hr^B \dpe\hp\dpe\hr^B}\\
&=0
\label{harmz}
\end{align}
where $\mathcal{D}$ is the cross-sectional disc domain, and $\partial\mathcal{D}$ is its boundary. 
This implies that $\hr^B$ is constant on $\mathcal{D}$, and therefore
vanishes. A similar argument applies to $\hr^\Phi$.
Furthermore, using (\ref{fg2}, \ref{fg3}) it follows
that $\{\hr^b, \hr^\varphi\}$ are constant on $\mathcal{D}$
and also may be chosen to vanish. Hence, (\ref{fg4}) and (\ref{fg5}) become
\begin{align}
&\cp\dpe W -2\partial^2_{\zeta u} W -\partial^2_{\zeta
    \zeta}W=P\rtu \label{wnotaper}\\
&\cp\dpe X -2\partial^2_{\zeta u} X - \partial^2_{\zeta\zeta} X
  =0\label{anotaper}
\end{align}
and are to be solved subject to the Dirichlet boundary condition at $r=R$ for $W$
and the Neumann boundary condition at $r=R$ for $X$. For the present
purposes, we are not interested in the source-free modes of the
waveguide, so we set $X=0$. Furthermore, charge conservation implies
that $\rho$ is independent of $\zeta$ and we expect the fields to
be independent of $\zeta$.


For a source given by 
\begin{equation}
\rho = f(u)\hat{\rho}(r,\theta)
\end{equation}
the $0$-form $P$ may be written
\[ P\rtu = \Xi(u)\hat{\rho} \quad\textrm{where}\quad \frac{\ud}{\ud u}\Xi(u)=f(u)\]
and a solution to (\ref{wnotaper}) is 
\[W = \Xi(u)\hat{W}(r,\theta)\quad \textrm {where}\quad\cp\dpe\hat{W}(r,\theta) = \hat{\rho}(r,\theta)\]
for $\hat{W}$ obeying the Dirichlet boundary condition at $r=R$.  The
electromagnetic $2$-form $F$ is
\begin{align}
&F = f(u)\ud u \wedge \dpe \hat{W}
\end{align}
which is compatible with the usual assumption that the electromagnetic
field vanishes ahead of and behind an ultra-relativistic source with
compact support in $u$ (see, for example,~\cite{chao82,chaobk}).
\subsection{Waveguide with a gradually changing radius}\label{gradualsec}
In this section we develop asymptotic expansions of solutions to the
preceding equations in an axially symmetric waveguide whose cross-section is a
slowly-varying function of $z=\zeta$.

A waveguide is considered to be ``slowly varying'' in $\zeta$ if
$R(\zeta)=\Rep(\epsilon\zeta)$ where $\epsilon>0$ is a small
dimensionless parameter. Hence, the waveguide boundary is described by
\begin{equation}\label{gradu1}
r-\Rep(\epsilon \zeta) = 0.
\end{equation}
Introduce a ``slow'' longitudinal co-ordinate $s$, which is defined by
\begin{equation}
s=\epsilon\zeta. 
\end{equation}
Rewrite all potentials as functions of $s$, using the notation
\begin{align}
\chi(r,\theta,\epsilon\zeta,u) &= \check{\chi}(r,\theta,s,u)\\
\partial_\zeta \chi(r,\theta,\epsilon\zeta,u) &= \epsilon \partial_s
\check\chi(r,\theta,s,u). 
\end{align}
A prime on a function accented with a caron
denotes differentiation with respect to $s$.
For example
\begin{eqnarray}
\frac{\partial}{\partial\zeta} \Rep(\epsilon\zeta) = \epsilon \frac{\partial}{\partial {(\epsilon \zeta)}} \Rep(\epsilon \zeta) = \epsilon \Rhat'(s).\label{gradu4}
\end{eqnarray}
Our approximation scheme follows by writing
\begin{equation}
\check\chi = \sum_{n=0}^\infty{\epsilon^n \check\chi_n},
\end{equation}
where
\begin{equation}
\check\chi\in\left\{\check A,\check a,\check B,\check
b,\check\Psi,\check\Phi,\check W,\check
X,\check\hr^B,\check\hr^b,\check\hr^\Phi,\check\hr^\varphi\right\}.
\label{asymser}
\end{equation}
Substituting such series into
(\ref{fg1}-\ref{fg5}) and the boundary conditions
(\ref{finalbound}) we obtain PDEs at each order $n$. 

Since $\partial_\zeta \chi = \epsilon \check\chi'$, 
equations (\ref{fg1}-\ref{fg5}) yield a set of transverse
Laplace and transverse Poisson equations at every order $n$, and the boundary 
conditions on $\check\hr^B_n$ and $\check\hr^\Phi_n$ now depend on
$(n-1)$-order potentials. This leads to a straightforward procedure
for calculating the potentials order-by-order. For each $n$ we
\begin{enumerate}
\item
Calculate the harmonic potential $\chr^B_n$ by solving the 2-dimensional Laplace equation
\begin{equation}\label{hrbharm}
\cp\dpe\chr^B_n=0\end{equation}
subject to the boundary condition~\footnote{Throughout this section, we
  are dealing with the \emph{transverse} Laplacian.  When considering
  the boundary conditions, $s$ can thus be treated as a parameter.}
\begin{equation}\label{hrB}
\chr^B_n=\check W_{n-1}' \quad\textrm{at }r=\Rhat(s).
\end{equation}
\item
Solve~\footnote{As $\chr^B_n$ and $\chr^\Phi_n$ are harmonic (with
  respect to the $2$-dimensional Laplacian), the converse of Poincar\'e's
  Lemma guarantees that (\ref{btob}) may be solved for $\chr^b_n$ and
  (\ref{hrphiphi}) may be solved for $\chr^\varphi_n$. The potentials
  $\chr^b_n$ and $\chr^\varphi_n$  are determined up to
  functions of $s$ and $u$ chosen so that the source in
  (\ref{ara}) is compatible with the Neumann boundary condition (\ref{cX_n_boun}).}  
\begin{equation}\label{btob}
\dpe\chr^b_n = \partial_u\hp\dpe\chr^B_n
\end{equation}
for $\chr^b_n$.
\item
Calculate the harmonic potential $\chr^\Phi_n$ by solving the
2-dimensional Laplace equation
\begin{equation}\label{hrphilap}
\cp\dpe\chr^\Phi_n=0\end{equation}
subject to the boundary condition
\begin{equation}\label{phinb}
\chr^\Phi_n=-\Rhat'(s)\frac{1}{r}\partial_\theta \cX_{n-1} \quad\textrm{at }r=\Rhat(s).
\end{equation}
\item
Solve~\footnote{The existence of a solution to
  (\ref{hrphiphi}) is guaranteed as $\chr^\Phi_n$ is harmonic.}
\begin{equation}\label{hrphiphi}
\dpe\chr^\varphi_n = \hp\dpe\chr^\Phi_n
\end{equation}
for $\chr^\varphi_n$.
\item
Calculate the potential $W_n$ by solving the 2-dimensional Poisson equation
\begin{align}
\label{wra}
\cp\dpe \cW_n= \cW_{n-2}'' +2\partial_{u}\cW_{n-1}'-\partial_u
\chr^\Phi_n-\partial_u \chr^B_n
- {\chr^B_{n-1}}'+ P_n
\end{align}
subject to the Dirichlet boundary condition
\begin{equation}
\cW_n = 0 \quad\textrm{at }r=\Rhat(s)
\end{equation}
where
\begin{equation}
P_n\rtu = 
\begin{cases} 
P\rtu\,\,&\text{for $n=0$}\\
0\,\,&\text{for $n\neq 0$}
\end{cases}
\end{equation}
and $\partial_u P\rtu=\rho\rtu$.
\item
Calculate the potential $\cX_n$ by solving the 2-dimensional Poisson equation
\begin{equation}
\cp\dpe \cX_n=\cX_{n-2}''+2\partial_{u} \cX_{n-1}'  -\chr^b_n-\partial_u \chr^\varphi_n-{\chr^\varphi_{n-1}}' \label{ara}
\end{equation}
subject to the Neumann boundary condition
\begin{equation}
\partial_r \cX_n=0\quad\textrm{at }r=\Rhat(s).\label{cX_n_boun}
\end{equation}
\end{enumerate}
We construct $\check A_n$, $\check a_n$, $\check B_n$, $\check b_n$,
$\check \Phi_n$ and $\check \Psi_n$ from 
\begin{align}
\check A_n &= \partial_u \check W_n + {\check W_{n-1}}'-\chr^B_n,\label{An}\\
\check B_n &= \chr^B_n-{\cW_{n-1}}',\label{Bn}\\
\check \Phi_n &=  -\cW_{n-2}'' -2\partial_{u}\cW_{n-1}'+\partial_u
  \chr^\Phi_n +\partial_u \chr^B_n + {\chr^B_{n-1}}'\label{Phin}\\
&= P_n\rtu-\cp\dpe \cW_n,\label{Phialternative}\\
\check a_n &= \partial_u \cX_n,\\
\check b_n &= \cX_{n-1}'-\chr^\varphi_n,\\
\check \Psi_n &= -\cX_{n-2}''-2\partial_{u} \cX_{n-1}',
  +\chr^b_n+\partial_u \chr^\varphi_n
+{\chr^\varphi_{n-1}}'\\
&= -\cp\dpe \cX_n.
\end{align}
Finally, we perform the summations to the required order and change
variable from $s$ to $\zeta$:
\begin{align}
&\check\chi(r,\theta,s,u)=\chi(r,\theta,\epsilon\zeta,u),\\
\label{stozeta}
&\check\chi'(r,\theta,s,u)=\epsilon^{-1}
\partial_\zeta\chi(r,\theta,\epsilon\zeta,u). 
\end{align}
Equation (\ref{F_in_terms_of_A})
then yields an asymptotic approximation for the electromagnetic
$2$-form $F$ with $\epsilon\ll 1$.
\subsection{Sources axially symmetric with respect to the waveguide}\label{axial}
When the source has rotational symmetry about the axis of the
waveguide, the fields sought are independent of
$\theta$ and determining auxiliary potentials independent of $\theta$ is straightforward. 

For each order $n$:
\begin{enumerate}
\item
$\chr^B_n$ is constant in the transverse waveguide cross-section, and is equal to the boundary value of $W_{n-1}'$:
\begin{equation}\label{hbaxi}
\chr^B_n=\cW_{n-1}'\vert_{r=\Rhat(s)}.
\end{equation}
\item
The other harmonic functions, and $\cX_n$ are zero,
\begin{equation}
\cX_n=\chr_n^b=\chr_v^\Phi=\chr_n^\varphi=0
\end{equation}
and hence
\begin{equation}
\check{a}_n=\check{b}_n=\check{\Psi}_n=0.
\end{equation}
\item
$\cW_n$ is calculated via the integral 
\begin{equation}
\cW_n = \int_{\Rhat(s)}^{r}{  \left[\int_{0}^{r_2}{\check\Omega_n
	r_1 \ud r_1}\right] \frac{\ud r_2}{r_2}}\label{wintaxi}
\end{equation}
where
\begin{equation}
\check\Omega_n=- \cW_{n-2}'' -2\partial_{u}\cW_{n-1}'+\partial_u
\chr^B_n + {\chr^B_{n-1}}'-P_n.\nn\\\label{axiomega}
\end{equation}
\item
Finally, $\{\check{A}_n$, $\check{B}_n$, $\check{\Phi}_n\}$ are 
\begin{eqnarray}
\check{A}_n &=& \partial_u \cW_n + {\cW_{n-1}}'-\chr^B_n,\label{Ans}\\
\check{B}_n &=& \chr^B_n-{\cW_{n-1}}',\label{Bns}\\
\check\Phi_n &=&  -\cW_{n-2}'' -2\partial_{u}\cW_{n-1}'+\partial_u
\chr^B_n + {\chr^B_{n-1}}'\\
&=&\check\Omega_n+P_n.\label{Phins}\end{eqnarray}
\end{enumerate}
For example,
\begin{align}
\check{\Phi}_0 =& 0,\\
\chr^B_0 =& 0,\\
\cW_0 =& -\frac{\Xi(u)}{2\pi}\left(\ln r - \ln \Rhat(s)\right),\label{w0axi1}\\
\chr^B_1 =& \frac{\Xi(u)}{2\pi}\frac{\Rhat'(s)}{\Rhat(s)},\\
\check\Phi_1 =& -\frac{f(u)}{2\pi}\frac{\Rhat'(s)}{\Rhat(s)},\label{phiax11}\\
\cW_1 =&
-\frac{f(u)}{8\pi}\frac{\Rhat'(s)}{\Rhat(s)}\left(r^2-\Rhat(s)^2\right),
\end{align}
\begin{align}
\chr^B_2 =& \frac{f(u)}{4\pi} \Rhat'(s)^2, \label{hb2ax1}\\
\check\Phi_2 =&
\frac{f'(u)}{4\pi}\left\{\left(\frac{\Rhat'(s)}{\Rhat(s)}\left(r^2-\Rhat(s)^2\right)\right)'+\Rhat'(s)^2\right\},\label{phi2ai1}\\
\cW_2 =&
\frac{f'(u)}{64\pi}\left(\Rhat(s)^2-r^2\right)\bigg\{\left(\frac{\Rhat'(s)}{\Rhat(s)}\right)'\left(3\Rhat(s)^2-r^2\right)
+ 4 \Rhat'(s)^2\bigg\}\label{axw2}
\end{align}
where $f'(u) = df(u)/du$.

The longitudinal impedance $Z^\parallel$ is
\begin{align}
\notag
Z^\parallel(\omega) &= -\frac{1}{I_\omega}\int^{\infty}_{-\infty}e^{-\frac{i\omega u}{c}}\Phi(r,\zeta,u)\,d\zeta\\
&= -\frac{1}{I_\omega}\sum_{n=0}^\infty \epsilon^{n-1}
\int^\infty_{-\infty}e^{-\frac{i\omega u}{c}}\check\Phi_n(r,s,u)\,ds
\label{on-axis_impedance_expansion}
\end{align}
and employing the above iterative procedure with the harmonic profile
\begin{equation}
\label{harm_profile}
f(u) =\frac{I_\omega}{\varepsilon_0 c} e^{\frac{i\omega u}{c}}
\end{equation}
leads to
\begin{align}
\label{on_axis_Phi0}
\check{\Phi}_0 =& 0,\\
\label{on_axis_Phi1}
\check{\Phi}_1 =& -\frac{e^{\frac{i\omega u}{c}}}{2\pi\varepsilon_0 c}\{\ln[\check{R}(s)]\}',\\
\label{on_axis_Phi2}
\check{\Phi}_2 =& 
  \frac{i\omega e^{i\omega
  u}}{4\pi\varepsilon_0 c^2}\left\{\left[\frac{\check{R}'(s)}{\check{R}(s)}\left(r^2-\check{R}(s)^2\right)\right]'
  +\check{R}'(s)^2\right\},\\
\notag
\check\Phi_3
=& \frac{e^{i\omega
    u}}{8\pi\varepsilon_0 c}\left\{\left(\frac{\Rhat'(s)}{\Rhat(s)}\right)'\left(\Rhat(s)^2-r^2\right)\right\}'
-\frac{\omega^2 e^{i\omega
    u}}{32\pi\varepsilon_0 c^3}\left(\Rhat(s)^2 \Rhat'(s)^2\right)'\\
&\label{on_axis_Phi3}
+\frac{\omega^2 e^{i\omega
    u}}{32\pi\varepsilon_0 c^3} \bigg\{\left(\Rhat(s)^2-r^2\right)\bigg[\left(\frac{\Rhat'(s)}{\Rhat(s)}\right)'\left(3\Rhat(s)^2-r^2\right)+4\Rhat'(s)^2\bigg]\bigg\}'.
\end{align}
Hence, introducing $Z^\parallel_{n\,\text{on-axis}}$ where
\begin{equation}
\label{on-axis_impedance_expansion_terms}
Z^\parallel_{n\,\text{on-axis}}(\omega) = -\frac{1}{I_\omega}\epsilon^{n-1} \int^\infty_{-\infty} e^{-i\omega u}
  \check\Phi_{n}(r,s,u)\, ds,
\end{equation}
it follows (\ref{on_axis_Phi1}, \ref{on_axis_Phi2})
yield (\ref{imp_Z0}, \ref{imp_Z2}). Clearly, $Z^\parallel_{0\,\text{on-axis}} =
0$ and the third order contribution (\ref{on_axis_Phi3}) to
$\check{\Phi}$ leads to $Z^\parallel_{3\,\text{on-axis}}= 0$ since we choose $\check R'(\infty)=\check R'(-\infty)=0$. 

Although expressions for $\check{\Phi}_n$ rapidly increase in
complexity with increasing $n$, they follow directly using the above
iterative procedure and are straightforward to generate using a
computer algebra package such as Maple~\cite{maple}. It may be shown
that $\{Z^\parallel_{5\,\text{on-axis}}, Z^\parallel_{7\,\text{on-axis}}\}$ vanish because
$\Rhat'(\infty)=\Rhat'(-\infty)=0$. Explicit expressions for
$\{Z^\parallel_{1\,\text{on-axis}},Z^\parallel_{2\,\text{on-axis}},Z^\parallel_{4\,\text{on-axis}},Z^\parallel_{6\,\text{on-axis}}\}$
were given in Section \ref{section:overview_and_results}.
\subsection{Source Offset from the Waveguide's Central Axis}
The general scheme developed in Section
\ref{gradualsec} simplifies when used to calculate the
geometric impedance of an infinitesimally thin beam offset by a
displacement $r_0$ from the central axis $r=0$:
\begin{equation}
\rho\rtu = f(u)\delta\left(x-r_0 \cos \theta_0)\right)\delta\left(y-r_0\sin\theta_0\right).
\end{equation}
We first calculate the zero order fields. As before, since $\chr^B_0$ and
$\chr^\Phi_0$ are harmonic and vanish on the waveguide boundary, it
follows $\chr^B_0 = \chr^\Phi_0 = 0$ and we choose
$\chr^b_0=\chr^\varphi_0=0$. Equation (\ref{wra}) gives
\begin{equation}\label{cdcw0}
\cp\dpe \cW_0 = \Xi(u)\delta\left(x-r_0 \cos \theta_0)\right)\delta\left(y-r_0\sin\theta_0\right)
\end{equation}
and a solution to (\ref{cdcw0}) which vanishes at $r=\Rhat(s)$ is
\begin{align}
\label{A0}
\cW_0 =&\frac{\Xi(u)}{4\pi}\bigg\{\ln\left(\frac{r^2 r_0^2}{{\Rhat(s)}^2} +
   {\Rhat(s)}^2 - 2 r r_0\cos(\theta-\theta_0)\right)
-\ln\left(r^2+r_0^2-2rr_0\cos(\theta-\theta_0)\right) \bigg\}.
\end{align}
It follows
\begin{align}
\cW_0=\frac{\Xi(u)}{4\pi}\bigg\{2\ln \Rhat(s) +\ln\left[ 1-\frac{r_0 r}{\Rhat(s)^2}
  e^{-i(\theta-\theta_0)}\right]
+\ln\left[ 1-\frac{r_0 r}{\Rhat(s)^2} e^{i(\theta-\theta_0)}\right]
+\dots\bigg\} 
\end{align}
where $\dots$ indicates terms independent of $s$.
For $r_0 < \Rhat(s)$ we have
\[
\left\vert \frac{r_0 r}{\Rhat(s)^2} e^{\pm i (\theta-\theta_0)}\right\vert < 1
\]
and using
\[
\ln(1-x) = -\sum_{m=1}^{\infty} \frac{x^m}{m} \quad\textrm{for}\,\, \vert x\vert<1
\]
it follows
\begin{align}
\cW_0 = \frac{\Xi(u)}{2\pi}\bigg\{&\ln \Rhat(s) - \sum_{m=1}^{\infty}
\frac{1}{m}\left(\frac{r_0 r}{\Rhat(s)^2}\right)^m \cos m(\theta-\theta_0)
+\dots\bigg\}
\end{align}
where $\dots$ indicates terms independent of $s$.
Differentiating $\cW_0$ with respect to $s$ yields 
\begin{align}
\label{w0primed}
\cW_0' = &\frac{\Xi(u)}{2\pi}\bigg\{\frac{\Rhat'(s)}{\Rhat(s)}
+
2\frac{\Rhat'(s)}{\Rhat(s)} \sum_{m=1}^\infty \left(\frac{r_0
  r}{\Rhat(s)^2}\right)^m \cos m(\theta-\theta_0)\bigg\}.
\end{align}
The first term in (\ref{w0primed}) is the contribution from the
monopole ($m=0$) term in the source and has the same form as $\cW_0'$
due to an on-axis thin beam. Thus, the monopole contribution has
already been discussed in Section \ref{section:impedance_on-axis_beam}
and here we concentrate on the terms that arise for $m>0$:  
\begin{equation}
\cW_{0,m}' = \frac{\Xi(u)}{\pi} \frac{\Rhat'(s)}{\Rhat(s)^{2m+1}} r^m
r_0^m \cos m(\theta-\theta_0).
\end{equation}
The final potentials and impedances can then be evaluated by summing
over $m$. A commonly used approximation for fields with $r,r_0 \ll
\Rhat(s)$ is to truncate the multipole series at the dipole 
contribution ($m=1$); see, for example,~\cite{stupakov07}.
For each order $n>0$, the 6-step procedure of Section \ref{gradualsec}
must be followed to obtain the auxiliary potentials. In practice, this
procedure is simple to implement for an infinitesimally thin beam and the main
features of the calculation are summarized below. 

Steps 1 and 3 are satisfied by
$\check{\hr}^B_n$ and $\check{\hr}^\Phi_n$ of the form
$h^{B,\Phi}(s,u) r^m\cos m(\theta-\theta_0)$ with $h^{B,\Phi}(s,u)$ determined by
a boundary condition. For steps 2 and 4, we can then choose
$\check{\hr}^{b}=\partial_u h^{B}(s,u) r^m\sin m(\theta-\theta_0)$ and
$\check{\hr}^{\varphi}= h^{\Phi}(s,u) r^m\sin m(\theta-\theta_0)$. 
The Poisson equations in steps 5 and 6 can always be written in the form
\begin{align}
\cp\dpe \check{W}_{n,m} = \sum\limits_{p=0}^{n-1} \kappa_{p,n,m}(s,u) r^{2p} r^m
  \cos m(\theta-\theta_0),
\end{align}
with $W_n=0|_{r=\check{R}(s)}$ and
\begin{equation}
\cp\dpe \check{X}_{n,m} = \sum\limits_{p=0}^{n-1} \tau_{p,n,m}(s,u) r^p
  r^m,
\end{equation}
with $\partial_r\check{X}_n=0|_{r=\check{R}(s)}$, where the $0$-forms
$\kappa_{p,n,m}(s,u)$ and $\tau_{p,n,m}(s,u)$ are known functions arising from previous iterations. Solutions to the above are
\begin{align}
&\check{W}_{n,m} =\sum\limits_{p=0}^{n-1}
\frac{\check{R}(s)^{2(p+1)}-r^{2(p+1)}}{4(p+1)(p+1+m)}\kappa_{p,n,m}(s,u)
\,r^m \cos m(\theta-\theta_0),\\
&\check{X}_{n,m}=\sum\limits_{p=0}^{n-1}\left[
  \frac{2(p+1)+m}{m}\check{R}(s)^{2(p+1)} -r^{2(p+1)}\right]
\frac{\tau_{p,n,m}(s,u)}{4(p+1)(p+1+m)}
r^m \sin m(\theta-\theta_0).
\end{align}
For example, this procedure leads to
\begin{align}
&\chr^B_{1,m} = \frac{\Xi(u)}{\pi} \frac{\Rhat'(s)}{\Rhat(s)^{2m+1}} r^m
r_0^m \cos m(\theta-\theta_0),\\
&\chr^\Phi_1 = 0,\\
\label{w1km}
&\cW_{1,m} = \frac{f(u)}{4\pi(1+m)} \frac{\Rhat'(s)}{\Rhat(s)^{2m+1}}
\left(\Rhat(s)^2-r^2\right) r^m r_0^m \cos m(\theta-\theta_0),\\
&\cX_{1,m} = -\frac{f(u)}{4\pi(1+m)}
\frac{\Rhat'(s)}{\Rhat(s)^{2m+1}}
\left[ \left(\frac{m+2}{m}\right)\Rhat(s)^2-r^2\right] r^m
r_0^m\sin m(\theta-\theta_0),\label{x1km}
\end{align}
\begin{align}
&\chr^B_{2,m} =  \frac{2f(u)}{4\pi(1+m)} \frac{\Rhat'(s)^2}{\Rhat(s)^{2m}}  r^m r_0^m \cos m(\theta-\theta_0),\label{hrB2km}\\
&\chr^\Phi_{2,m} =  \frac{2f(u)}{4\pi(1+m)}
\frac{\Rhat'(s)^2}{\Rhat(s)^{2m}}  r^m r_0^m \cos m(\theta-\theta_0),
\label{hrP2km}\\
&\cW_{2,m} = \frac{f'(u)}{8\pi(1+m)} \left(
\frac{\Rhat'(s)}{\Rhat(s)^{2m+1}} \right)'
\left[\Rhat(s)^2\frac{\Rhat(s)^2-r^2}{1+m}-\frac{\Rhat(s)^4-r^4}{2(2+m)}\right]
r^m r_0^m \cos m(\theta-\theta_0), \label{w2stm}\\
&\cX_{2,m} =\Bigg\{ \left(\frac{\Rhat'(s)}{\Rhat(s)^{2m+1}}\right)'\bigg[
  \frac{\frac{m+4}{m}\Rhat(s)^4-r^4}{4(m+2)}
+\frac{m+2}{2m(m+1)}\Rhat(s)^2\left(r^2-\frac{m+2}{m}\Rhat(s)^2\right)\bigg]\\
&\qquad\qquad +\frac{\Rhat'(s)^2}{\Rhat(s)^2}\frac{2}{m}\left(r^2-\frac{m+2}{m}\Rhat(s)^2\right)\Bigg\}
\frac{f'(u)}{4\pi(m+1)}r_0^m r^m\sin m(\theta-\theta_0)
\end{align}
where $f'(u)=df(u)/du$.

Furthermore,
it follows
\begin{align}
&\check\Phi_{0,m} = 0,\\
&\check\Phi_{1,m}
= \frac{2}{m} \frac{f(u)}{4\pi} \left(\frac{1}{\Rhat(s)^{2m}}\right)'
r^m r_0^m \cos m(\theta-\theta_0),\\
\check\Phi_{2,m} =&
\frac{f(u)}{4\pi(1+m)}\bigg\{2\left(\frac{\Rhat'(s)}{\Rhat(s)^{2m+1}}(r^2-\Rhat(s)^2)\right)'
+4 \frac{\Rhat'(s)^2}{\Rhat(s)^{2m}}\bigg\} r^m r_0^m \cos m(\theta-\theta_0).
\end{align}

The multipole impedances introduced in Section
\ref{section:overview_and_results} are obtained by developing the
above to higher order in $n$ using (\ref{harm_profile},~\ref{on-axis_impedance_expansion}).
\section{Acknowledgements}
We thank the Cockcroft Institute for support.
\appendix
\section{Decomposition of 1-forms in 2-dimensional, Simply Connected,
  Bounded Domains}\label{decompose}
The method used here to analyse Maxwell's equations requires the Hodge
decomposition of forms on a manifold with boundary~\cite{schwarz}.

Let $\doma$ be a transverse cross-section of an axially symmetric waveguide.  We
have the immersion map $\iota:\domb\rightarrow \doma$, so that
$\iota^*\alpha$ is the pull-back of a $1$-form $\alpha$ onto the
boundary $\domb$ of $\doma$. The set of smooth 1-forms on $\doma$ can be
decomposed as:
\begin{equation}\label{decomp1}
\Lambda_1(\doma) = \dpe \matf_d(\doma) \oplus \hp\dpe\matf(\doma)\end{equation}
where $\matf(\doma)$ is the space of smooth 0-forms on $\doma$ and
$\matf_d(\doma)$ is the subspace of $\matf(\doma)$ whose elements
satisfy the Dirichlet boundary condition:
\begin{equation}
\matf_d(\doma) = \left\{f\in\matf_d(\doma) \mid \iota^*f=0\right\}.
\end{equation}

To demonstrate the Hodge decomposition (\ref{decomp1}) we show
$\dpe\matf_d(\doma)$ and $\hp \dpe\matf(\doma)$ are orthogonal with
respect to the symmetric product:
\begin{equation}
\label{inner_product}
(\mu,\nu) := \int_\doma{\mu \wedge\hp\nu}
\end{equation}
where $\mu$ and $\nu$ are 1-forms on $\doma$.  We will then show that if a 1-form is orthogonal to
$\dpe\matf_d(\doma)$ then that 1-form must be in  $\hp\dpe\matf(\doma)$.

For $A \in \matf_d(\doma)$ and $a \in \matf(\doma)$
\begin{align}
\notag
(\dpe A, \hp\dpe a) &= \int_\doma{\dpe A \wedge\hp(\hp \dpe a)}\\
\notag
&= -\int_\doma{\dpe A \wedge \dpe a}\\
&= -\int_\domb{\iota^*(A\dpe a)}=0
\end{align}
and we conclude that $\dpe\matf_d(\doma)$ and $\hp \dpe\matf(\doma)$
are orthogonal with respect to (\ref{inner_product}).

Let $\omega$ be a $1$-form orthogonal to $\dpe\matf_d(\doma)$.
Thus, for every $A\in\matf_d(\doma)$ we have
\begin{equation}
0=(\dpe A,\omega)=\int_\doma{\dpe A\wedge \hp\omega}.
\end{equation}
However, as $A$ satisfies the Dirichlet boundary condition, we have
\begin{align}
\notag
0 &= \int_\domb{\iota^*(A \hp\omega)}\\
\notag
 &= \int_{\doma}{\left(\dpe A \wedge \hp\omega+A \dpe\hp\omega\right)}\\
&= \int_{\doma}{A \dpe\hp\omega}
\end{align}
and, as this is true for every $A\in\matf_d(\doma)$, we conclude
\begin{equation}
\dpe\hp\omega=0.
\end{equation}
The cross-section ${\cal D}$ of the axially symmetric
waveguide considered here is simply connected. By the converse of
Poincar\'e's Lemma~\cite{abraham}, we can therefore write
\begin{equation}
\hp\omega=\dpe \kappa \Rightarrow \omega=-\hp\dpe\kappa
\end{equation}
which is clearly a member of $\hp \dpe \matf(\doma)$.

Thus,
any 1-form $\alpha$ on $\doma$ may be written as
\begin{equation}
\label{Hodge_deRham_decomp}
\alpha = \dpe A + \hp\dpe a
\end{equation}
for some $A\in\matf_d(\doma)$ and some $a\in\matf(\doma)$.
\section{Motivation behind auxiliary
  potentials}\label{juste}
\subsection{Maxwell Equations}
The approach used here to decompose Maxwell's equations employs the
Hodge decomposition (\ref{Hodge_deRham_decomp}) on transverse
$1$-forms. In earlier Sections we showed that the
introduction of auxiliary potentials and our approximation method lead to Poisson and Laplace
equations for the auxiliary potentials.

Alternatively, one may try to directly generalize the approach
in~\cite{stupakov07} by writing the Maxwell equations in terms of the
electric field and postulating asymptotic series for $A,a$. However,
as we will now show, the latter approach leads to third order PDEs that are
harder to analyse than the hierarchy of Poisson and Laplace equations
obtained earlier for the auxiliary potentials $\{\mathcal{H}^\Phi,
\mathcal{H}^\varphi, \mathcal{H}^B,
\mathcal{H}^b, W,X\}$.

The electric field $1$-form $E$ is
\begin{equation}
E
= \Phi \ud\zeta - \alpha
\end{equation}
where $\Phi$ is the longitudinal component of the electric field and
$-\alpha$ is the transverse projection of $E$.
The Maxwell system (\ref{maxu1}-\ref{maxu6}) can used to obtain
equations for the electric field alone:
\begin{align}
&\cp\alpha = \rho-\partial_u\Phi -\partial_\zeta \Phi,\label{appb1}\\
&\cp\dpe\Phi= 2\partial^2_{\zeta u}\Phi+
  \partial^2_{\zeta\zeta}\Phi,\label{appb2}\\
&\cp\dpe\alpha + \dpe\cp\alpha = \dpe\rho + 2\partial^2_{\zeta u}\alpha+\partial^2_{\zeta\zeta}\alpha.\label{appb3}
\end{align}
\subsection{Boundary Conditions}
The boundary of the waveguide is perfectly conducting so
the electric field tangent to the boundary vanishes:
\begin{equation}
\label{E_boundary_condition}
E\wedge \ud{\cal S} = 0|_{r=R(\zeta)}
\end{equation}
where
\begin{equation}
{\cal S}=r-R(\zeta).
\end{equation}
The boundary condition (\ref{E_boundary_condition}) with 
\begin{equation}
\alpha=\alpha_r\ud r+\alpha_\theta \ud\theta
\end{equation}
leads to
\begin{align}
&\alpha_\theta=0|_{r=R(\zeta)},\label{appbb1}\\
&\Phi-R'(\zeta)\alpha_r=0|_{r=R(\zeta)}.\label{appbb2}
\end{align}
\subsection{Hierarchy of approximations}
Inserting the Hodge-de Rham decomposition of $\alpha$
\begin{equation}
\alpha = \dpe A + \hp\dpe a
\end{equation}
into (\ref{appb1}-\ref{appb3}), where $A=0|_{r=R(\zeta)}$, yields
\begin{align}
&\cp\dpe A = \rho - \partial_u\Phi - \partial_\zeta \Phi,\\
&\cp\dpe\Phi= 2\partial^2_{\zeta u}\Phi+ \partial^2_{\zeta\zeta}\Phi,\\
&\hp\dpe\cp\dpe a + \dpe\cp\dpe A = \dpe\rho
+ \left(2\partial^2_{\zeta u}+\partial^2_{\zeta\zeta}\right)\left(  \dpe A + \hp\dpe a \right).
\end{align}
As $A$ satisfies the Dirichlet boundary condition, (\ref{appbb1}) and
(\ref{appbb2}) become
\begin{align}
&\partial_r a=0|_{r=R(\zeta)},\\
&\Phi+R'(\zeta)\left(\frac{1}{r}\partial_\theta a - \partial_r
  A\right)=0|_{r=R(\zeta)}.
\end{align}
Introducing an asymptotic gradual-taper expansion in the same way as
in the main body of the text gives
\begin{align}
&\cp\dpe \check{A}_0=\rho, 
&\check{\Phi}_0=\check{a}_0=0
\end{align}
with $\check{A}_0=0|_{r=\check{R}(s)}$ and, for $n>0$, $\check{\Phi}_n$ satisfies
\begin{equation}\label{phiapprr}
\cp\dpe\check{\Phi}_n = -2\partial_u\check{\Phi}_{n-1}'-\check{\Phi}_{n-2}''
\end{equation}
subject to the boundary condition
\begin{equation}
\check{\Phi}_n = \check{R}'(s)\left(\partial_r
\check{A}_{n-1}-\frac{1}{r}\partial_\theta \check{a}_{n-1}\right)\bigg|_{r=\check{R}(s)}. 
\end{equation}
Then, $\check{A}_n$ is calculated from 
\begin{equation}\label{lapan}
\cp\dpe \check{A}_n = -\partial_u\check{\Phi}_n-\check{\Phi}_{n-1}'
\end{equation}
subject to the Dirichlet boundary condition $\check{A}_n=0|_{r=\check{R}(s)}$.
Finally, $\check{a}_n$ is calculated from 
\begin{align}
\label{3opde2}
\hp\dpe\cp\dpe \check{a}_n = \partial_u \dpe \check{\Phi}_{n} + \dpe \check{\Phi}_{n-1}'+
2\partial_u\dpe \check{A}_{n-1}'
+ 2\partial_u \hp\dpe \check{a}_{n-1}' + \dpe \check{A}_{n-2}''+ \hp\dpe \check{a}_{n-2}''
\end{align}
with $\partial_r \check{a}_n=0|_{r=\check{R}(s)}$.
Writing (\ref{3opde2}) in $(r,\theta)$ components yields a coupled pair of third
order PDEs which is more difficult to tackle, in
general, than the sequence of Laplace and Poisson equations for
auxiliary potentials found earlier. However, if one can find functions $\check{\phi}_n$
and $\check{\psi}_n$ such that 
\begin{equation}\label{psiphinonhar}
\dpe\check{\phi}_n = -\hp\dpe\check{\Phi}_n, \qquad \dpe \check{\psi}_n = -\hp\dpe \check{A}_n
\end{equation}
then (\ref{3opde2}) can be converted to a two-dimensional Poisson
equation (see equations (36, 37) in \cite{stupakov07} for a
similar approach):
\begin{align}
\cp\dpe\check{a}_n =& \partial_u\check{\phi}_{n} + \check{\phi}_{n-1}'+
2\partial_u\check{\psi}_{n-1}' + 2\partial_u \check{a}_{n-1}
+ \check{\psi}_{n-2}''+ \check{a}_{n-2}''.
\end{align}
Unfortunately, applying the transverse exterior derivative $\ud_\perp$ to
(\ref{psiphinonhar}) gives non-trivial conditions on $\check{\Phi}_n$ and $\check{A}_n$: 
\begin{equation}
0=\cp\dpe \check{\Phi}_n, \qquad 0=\cp\dpe \check{A}_n.
\end{equation}
This implies that we can only find functions $\check{\phi}_n$ and
$\check{\psi}_n$ that satisfy (\ref{psiphinonhar}) if $\check{\Phi}_n$
and $\check{A}_n$ are harmonic.  
Thus, solutions to (\ref{psiphinonhar}) will not exist if the right
hand sides of (\ref{phiapprr}) and (\ref{lapan}) are non-zero. If
the right hand side of (\ref{lapan}) is zero it
follows $\check{A}_n=0$ since $\check{A}_n$ satisfies Laplace's
equation in ${\cal D}$ and $\check{A}_n = 0|_{r=\check{R}(s)}$.
Thus, only in rare cases can we avoid having to solve
third order PDEs. One way to avoid this problem is to employ the
methods presented in the main text.

\end{document}